\documentclass[letterpaper]{article} 
\usepackage{aaai25}  
\usepackage{times}  
\usepackage{helvet}  
\usepackage{courier}  
\usepackage[hyphens]{url}  
\usepackage{graphicx} 
\usepackage{natbib}  
\usepackage{caption} 
\usepackage{algorithm}
\usepackage{algorithmic}

\usepackage{color}
\usepackage{comment}
\usepackage{listings}
\usepackage{booktabs}
\usepackage{multirow}

\usepackage{xcolor}

\usepackage{graphicx}
\usepackage{caption}
\usepackage{subcaption}

\definecolor{VulBinCode}{HTML}{CC6600}
\definecolor{DarkGreen}{HTML}{3C8031}

\usepackage{tikz}
\usepackage{amsmath}

\usepackage{filecontents}
\usepackage{comment}
\usepackage{booktabs}
\usepackage{algorithm}
\usepackage{algorithmic}
\usepackage{multirow}
\usepackage{listings}
\usepackage{xcolor}

\usepackage{newfloat}
\usepackage{listings}
\DeclareCaptionStyle{ruled}{labelfont=normalfont,labelsep=colon,strut=off} 
\floatstyle{ruled}
\newfloat{listing}{tb}{lst}{}
\floatname{listing}{Listing}
\title{Enhancing Reverse Engineering: Investigating and Benchmarking Large Language Models for Vulnerability Analysis in Decompiled Binaries}
\author{
    Dylan Manuel\textsuperscript{\rm 1}\textsuperscript{\rm 2},
    Nafis Tanveer Islam\textsuperscript{\rm 1}\textsuperscript{\rm 2},
    Joseph Khoury \textsuperscript{\rm 3}, 
    Ana Nunez\textsuperscript{\rm 1}\textsuperscript{\rm 2}, \\
    Elias Bou-Harb, \textsuperscript{\rm 3}
    Peyman Najafirad\textsuperscript{\rm 1}\textsuperscript{\rm 2}
}
\affiliations{
    \textsuperscript{\rm 1}Secure AI an Autonomy Laboratory \\
    \textsuperscript{\rm 2} University of Texas at San Antonio \\
    \textsuperscript{\rm 3} Louisiana State University

}

\usepackage{bibentry}

\begin{document}

\maketitle

\begin{abstract}
    Security experts reverse engineer (\textit{decompile}) binary code to identify critical security vulnerabilities. The limited access to source code in vital systems -- such as firmware, drivers, and proprietary software used in Critical Infrastructures (CI) -- makes this analysis even more crucial on the binary level. Even with available source code, a semantic gap persists after compilation between the source and the binary code executed by the processor. This gap may hinder the detection of vulnerabilities in source code. That being said, current research on Large Language Models (LLMs) overlooks the significance of decompiled binaries in this area by focusing solely on source code. In this work, we are the first to empirically uncover the substantial semantic limitations of \textit{state-of-the-art} LLMs when it comes to analyzing vulnerabilities in decompiled binaries, largely due to the absence of relevant datasets. To bridge the gap, we introduce \textcolor{VulBinCode}{\textbf{DeBinVul}}, a novel \underline{de}compiled \underline{bin}ary code \underline{vul}nerability dataset. Our dataset is multi-architecture and multi-optimization, focusing on C/C++ due to their wide usage in CI and association with numerous vulnerabilities. Specifically, we curate \textit{\textbf{150,872}} samples of vulnerable and non-vulnerable decompiled binary code for the task of \textbf{\textit{(i)}} identifying; \textbf{\textit{(ii)}} classifying; \textbf{\textit{(iii)}} describing vulnerabilities; and \textbf{\textit{(iv)}} recovering function names in the domain of decompiled binaries. Subsequently, we fine-tune \textit{state-of-the-art} LLMs using \textcolor{VulBinCode}{\textbf{DeBinVul}} and report on a performance increase of \textbf{\textit{19\%}}, \textbf{\textit{24\%}}, and \textbf{\textit{21\%}} in the capabilities of \texttt{CodeLlama}, \texttt{Llama3}, and \texttt{CodeGen2} respectively, in detecting binary code vulnerabilities. Additionally, using \textcolor{VulBinCode}{\textbf{DeBinVul}}, we report a high performance of \textbf{\textit{80-90\%}} on the vulnerability classification task. Furthermore, we report improved performance in function name recovery and vulnerability description tasks. \textit{All our artifacts are available at} \footnote{\url{https://anonymous.4open.science/r/vuln-decompiled-summarization-8017}}.\
    
    \end{abstract}

    \section{Introduction}

It is crucial to perform vulnerability analysis in software that plays a vital role in shaping Critical Infrastructure (CI) sectors such as water, energy, communications, and defense, to name a few. Despite the many advancement in software security, the reported number of Common Vulnerabilities and Exposures (CVEs) has been increasing annually, from \textit{\textbf{14,249}} in 2022, to \textit{\textbf{17,114}} in 2023, and surging to \textit{\textbf{22,254}} in 2024 \cite{qualys}. These CVEs are correlated with the Common Weakness Enumeration (CWE) categories maintained by \texttt{MITRE}, which provide a baseline for identifying, mitigating, and preventing security weaknesses during source code development. Notably, during the compilation optimization, the source code transitions into binary code, resulting in mismatches and changes in code properties \cite{eschweiler2016discovre}. This inherently creates a vulnerability semantic discrepancy not addressed by CWEs or other vulnerability categorization systems. As such, vulnerability analysis of source code and binary code remains two distinct and separate areas of research \cite{mantovani2022convergence}. This phenomenon is succinctly captured by the statement, \textbf{\textit{``What you see is not what you execute''}} \cite{balakrishnan2010wysinwyx}.\\


\noindent \textbf{Why Decompiled Binary Code Vulnerability Analysis? --- \textit{Significance \& Technical Challenges}.}
Binary code (i.e., binaries/executables) is a fundamental component of computing and digital systems taking the form of firmware, drivers/agents, and closed-source software. To safeguard these systems, reverse engineers attempt to uncover source code from binary code using decompilation tools such as \texttt{Ghidra}, \texttt{angr}, and \texttt{IDA Pro}, subsequently performing essential vulnerability analysis on decompiled binary code \cite{burk2022decomperson}. This is particularly important for two main reasons; first, access to source code is most of the time limited/restricted for proprietary or security reasons; second, vulnerabilities may not be apparent in the source code, such as those related to the execution environment, operating system, specific compiler optimizations, and hardware specifications. For instance, \texttt{use-after-free} (memory corruption) vulnerabilities, which affect many closed-source system components and network protocols written in \texttt{C/C++}, are known to be one of the most difficult types to identify using source code static analysis. \cite{lee2015preventing, nguyen2020binary}. 
On a different note, due to the NP-complete nature of the compiler optimization problem \cite{eschweiler2016discovre}, decompiled binary code loses important constructs, such as structured control flow, complex data structures, variable names, and function signatures \cite{burk2022decomperson}. As a consequence, these setbacks impede the ability of reverse engineers to analyze vulnerability in binary code, necessitating significant manual effort and time investment.\\






\noindent \textbf{Avant-garde Advancements and Perceived Opportunities.}
More recently, \textit{state-of-the-art} Large Language Models (LLMs) have been employed as an optimizer to improve the readability and simplicity of decompilers' output, ultimately reducing the cognitive burden of understanding decompiled binary code for reverse engineers \cite{hu2024degpt}. Similarly, a cross-modal knowledge prober coupled with LLMs have been utilized to effectively lift the semantic gap between source and binary code \cite{su2024source}. Furthermore, comprehensive benchmarking was conducted on \texttt{ChatGPT/GPT-4} and other LLMs to evaluate their effectiveness in summarizing the semantics of binary code  \cite{jin2023binary}. This assessment revealed the transformative capabilities of LLMs in the field while also highlighting key findings on their limitations, which demands further research. While these efforts aim to improve the readability and comprehension of decompiled binary code semantics, they overlook the vulnerability semantic gap between source code and deccompiled binary. To date, no comprehensive research has been conducted to thoroughly investigate and explore the potential of LLMs in decompiled binary code vulnerability analysis. This task remains far from straightforward due to the following two main limitations; \textbf{\textit{(i)}} \textit{\textbf{lack of real-world decompiled binary code vulnerability datasets; and}}  \textbf{\textit{(ii)}} \textbf{\textit{vulnerability semantic gap between source and decompiled binary code in LLMs.}} Currently \textit{state-of-the-art} LLMs are trained on textual-like input, including source code, but they lack semantic knowledge of vulnerabilities in the decompiled binary code domain due to the absence of representative datasets. Through an empirical and pragmatic investigation of the analytical abilities of LLMs, we find a consistent low performance of \textit{67\%}, \textit{54\%}, and \textit{33\%} in decompiled binary code, compared to a slightly higher performance of \textit{75\%}, \textit{68\%}, and \textit{45\%} in source code with \texttt{GPT4}, \texttt{Gemini}, and \texttt{LLaMa 3}, respectively. Table 1 highlights some of the insights we derived from our investigation. More information on the investigation is provided in the Appendix and Table \ref{tab:motivation} in Section Source \& Decompiled Binary Code Vulnerability Semantic Gap: \textit{Investigating LLMs' Analytical Abilities}. To this end, significant manual effort is required to curate decompiled binary code samples that include relevant vulnerabilities, realistic compilation and decompilation settings, and representative input formats for LLMs. Moreover, this entails of \textit{state-of-the-art} LLMs through extensive fine-tuning and instructive/prompting techniques.\\

\begin{table}[t]
\centering
\caption{\textbf{Motivational Investigation:} LLMs semantic gap comparison between static source code and decompiled binary code on vulnerability classification task. Reported average F1-scores.}
\begin{tabular}{@{}llll@{}}
\toprule \toprule
\textbf{Input Type} & \textbf{GPT4}    & \textbf{Gemini}  & \textbf{CodeLLaMa} \\ \midrule \midrule
\textbf{Source Code}     & 0.75            & 0.68             & 0.64               \\
\textbf{Dec. Binary Code} & 0.67 \textcolor{red}{$\downarrow$}            & 0.54 \textcolor{red}{$\downarrow$}            & 0.54 \textcolor{red}{$\downarrow$}              \\\\ \midrule\midrule 
                    & \textbf{Mistral} & \textbf{LLaMa 3} & \textbf{CodeGen2}  \\ \midrule \midrule
\textbf{Source Code}     & 0.60             & 0.45             & 0.64               \\
\textbf{Dec. Binary Code} & 0.54 \textcolor{red}{$\downarrow$}            & 0.33 \textcolor{red}{$\downarrow$}            & 0.52 \textcolor{red}{$\downarrow$}              \\ \bottomrule
\end{tabular}
\end{table}

\noindent \textbf{Our Contribution.} To tackle these challenges and capitalize on the perceived opportunities, this work aims to ask:

\begin{center}
    \textbf{\textit{Can we enhance reverse engineering by bridging the semantic gap between source and decompiled binary code vulnerability analysis in state-of-the-art LLMs?}}
\end{center}


\noindent To answer this question, we undertake the following quests. \textbf{\textit{Firstly,}} we empirically investigate the analytical abilities of \textit{state-of-the-art} LLMs and uncover a vulnerability semantic gap between source and decompiled binary code. Our investigation encompasses real-world code injection in public repositories, simulating an emergent cybersecurity attack that targets the widely recognized Linux-based \texttt{XZ Utils} \cite{akamai}. \textbf{\textit{Secondly,}} we introduce \textcolor{VulBinCode}{\textbf{DeBinVul}} a novel \underline{de}compiled \underline{bin}ary \underline{vul}nerability dataset with zero-shot prompt engineering. Our dataset comprises relevant non-vulnerable and vulnerable source code samples, tagged with CWE classes, and compiled using \texttt{Clang} and \texttt{GCC} across \textit{four} different CPU architectures: \texttt{x86}, \texttt{x64}, \texttt{ARM}, and \texttt{MIPS}. During compilation, we applied \textit{two} levels of optimizations; \texttt{$O_{0}$} and \texttt{$O_{3}$}. Then, using \texttt{GHIDRA} we decompile the compiled code to obtain the decompiled binary code samples. Furthermore, we augment our dataset with code descriptions and instruction/prompting techniques. \textbf{\textit{Thirdly,}} we fine tune and optimize \textit{state-of-the-art} LLMs, aiming to enhance their capabilities in assisting reverse engineers in uncovering vulnerabilities in decompiled binary code. In summary, the contributions of this paper are  as follows:

\begin{itemize}


    \item To the best of our knowledge, we are the first to empirically investigate the vulnerability semantic gap between source and decompiled binary code in \textit{state-of-the-art} LLMs. Our findings highlight the suboptimal performance of these models in performing vulnerability analysis on decompiled binaries.

    \item We compile and release, \textcolor{VulBinCode}{\textbf{DeBinVul}}, a novel \underline{de}compiled \underline{bin}ary code \underline{vul}nerability dataset comprising \textit{\textbf{150,872}} samples of \textit{openly sourced}, \textit{synthetically generated}, and \textit{manually crafted corner case} \texttt{C/C++} code samples. It is designed to tackle four important binary code vulnerability analysis tasks, including \textit{vulnerability detection}, \textit{classification}, \textit{description}, and \textit{function name recovery}.

    

    \item We employ our proposed dataset to fine-tune and enhance the reverse engineering capabilities across a range of \textit{state-of-the-art} LLMs. Our results shows a performance increase of \textbf{\textit{19\%}}, \textbf{\textit{24\%}}, and \textbf{\textit{21\%}} in the capabilities of \texttt{CodeLlama}, \texttt{Llama 3}, and \texttt{CodeGen2} respectively, in detecting vulnerabilities in binary code.

\end{itemize}


    \section{Proposed Approach}
    
\begin{figure*}[t]
    \centering
    \includegraphics[width=1\linewidth]{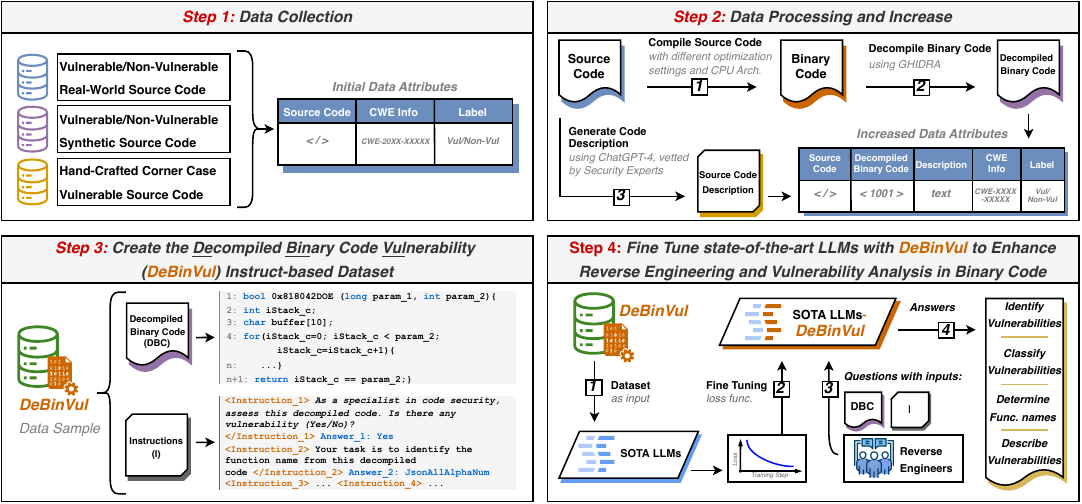}
    \caption{\textbf{Our Proposed Approach:} An overview of our proposed instruct dataset \textbf{\textcolor{VulBinCode}{DeBinVul}} with a sample example comprising a decompiled binary code input and a list of questions (instructions) and answers. Subsequently, using \textbf{\textcolor{VulBinCode}{DeBinVul}}, we train state-of-the-art LLM models to optimize them and elevate their capabilities in assisting reverse engineers in unveiling vulnerabilities in binary code.}
    \label{fig:arch}
\end{figure*}

In order to mitigate the challenges faced by LLMs in understanding decompiled binaries and improve their performance in understanding their security impact, we propose an extensive dataset comprised of source code and their decompiled binaries. Further details are provided in the sequel. Figure \ref{fig:arch} highlights our entire architecture.




\subsection{Step 1: Data Collection}


We compiled our dataset from three distinct sources: the National Vulnerability Database (NVD), the Software Assurance Reference Dataset (SARD), and a collection of real-world code enhanced with synthetically added vulnerabilities to cover corner cases where real-world code is not available for certain vulnerabilities for proprietary and security reasons.

\textbf{NVD.} NVD provides real-world source code vulnerabilities which were reported by developers and security experts from various repositories. But these are often individual functions, making it difficult to compile them into executables and decompile them due to unclear documentation and library dependencies. Therefore, during collection, we had to skip the ones which were not compilable.

Additionally, the NVD often lacks coverage of preparatory vulnerabilities, such as those involving specific configurations or security mechanism bypasses, which don't lead directly to exploits but can set the stage for more severe issues. While the information on the vulnerabilities is exposed, the source code is not exposed to NVD since it may contain sensitive information like code or data structure of the system, file, or operating system. These preparatory or indirect vulnerabilities are often not published in the NVD, as they require specific, often complex conditions to manifest in a real-world exploit.



\textbf{SARD.} SARD is a valuable resource for the software security community. It's a curated collection of programs containing deliberate vulnerabilities. Researchers and tool developers use SARD to benchmark their security analysis tools, identifying strengths and weaknesses. By exposing programs to a wide range of known vulnerabilities. SARD, while providing code examples with known vulnerabilities and all the code samples are executable, it often lacks real-world complexity and diversity.

\textbf{Vulnerability Injection.} Furthermore, we proposed an innovative automatic injection process to inject vulnerabilities in the source of real-world repositories to emulate this scenario. Table \ref{tab:test_models} briefly describes the various LLMs we analyze for vulnerability. After getting the repositories, we injected vulnerabilities into randomly selected functions from the repositories by prompt engineering using LLMs. We selected 8 of the top 25 CWE vulnerabilities from MITRE that are common in C/C++ programs. Out of the initial 500 randomly selected code samples, we injected vulnerabilities into 462, of which 38 were not compilable and were subsequently ignored. We provide the details of the repository selection and vulnerability injection process in the \textcolor{black}{Appendix \ref{appen:injection}(\textit{Vulnerability Injection Process})}.


Combining the SARD and the NVD for training LLMs can significantly enhance their capabilities in vulnerability analysis. While these two datasets offer either fully synthetic or fully real vulnerabilities, our method of injecting vulnerabilities tries to overcome the issues we see in NVD. Together, these datasets allow the LLM to generalize from structured, annotated examples to broader, complex, real-world scenarios, resulting in a more versatile model that can analyze decompiled binaries across various software contexts.


Hence, we opt to construct a dataset \textcolor{VulBinCode}{\textbf{DeBinVul}} by extending the capabilities of MVD to analyze decompiled binaries with instructions to align with LLMs.

\subsection{Step 2: Data Processing and Increase}
\textbf{Function Extraction.}
We developed a methodology using the Tree-Sitter parser to extract the functions from a file. Since we are extracting \texttt{C/C++} functions, we use \texttt{C} or \texttt{C++} grammar for function extraction. If the file name suffix is ``.c,'' then a C grammar is used; otherwise, if the suffix is ``.cpp,'' a \texttt{C++} grammar is used. The available functions from SARD have some special signatures in the function declaration which helps us to determine whether the function is vulnerable or not. For example, if the function name contains the term ``good'' or ``bad'', we consider them non-vulnerable or vulnerable functions. Furthermore, if the function if vulnerable, the function name also contains the CWE number as well. We utilize this information to annotate the vulnerable and the non-vulnerable functions and find the CWE numbers of the vulnerable function using regular expression. The code from NVD and our injection technique are functions. Therefore, they do not need to be extracted.


\textbf{Compilation and Optimization.} After we have extracted the vulnerable and non-vulnerable functions, we locate the necessary header files and other source code files needed to compile the CWE file. These files are conveniently located in the same directory as the CWE file. Each source code function was compiled six times to ensure comprehensive analysis, resulting in six binaries of a single function. This process involved using two compilers, two optimization levels, and four architectures.

\textbf{Decompilation.}
We used \texttt{Ghidra} \cite{ghidra} to decompile the functions. Decompilation can usually be done two ways, stripping and non-stripping the function or variable names. In real-world applications, the functions are variable names are stripped due to security reasons. Therefore, we emulate the same process by stripping function and variable names during compilation using a special flag $-s$ during compilation.


\textbf{Description.}
The functions we extracted mainly contain comments written by software security experts. However, these comments are partial and explain only a particular statement. However, there are multiple levels of comments for some important vulnerable-prone lines. Therefore, we use tree-sitter to create a method to define comments in C/C++. Then, we use the definition of the method to extract the comments inside these functions. Finally, we use the source code without the comments and the extracted comments and prompt GPT-4 to write a clean, comprehensive description of the code using a prompt. Furthermore, we also want to ensure that we use the same description when we use decompiled functions. Therefore, ensure that function and variable names are not present when describing the function objectives and vulnerabilities. 



\subsection{Step 3: Instructions}
We provide an instruction-based dataset, enabling the user or developer to use our system by providing instructions with code. Therefore, we created four types of robust instructions. We created four carefully curated prompts to instruct GPT-4 to create 20 instructions for each task; therefore, we have 80 instructions. Moreover, we provided 2 sample examples with each prompt that would guide GPT-4 to generate the most appropriate comment. These instructions are manually appended during training and testing with the input code based on the desired task. Table \ref{tab:prompt} shows the prompts we used to generate 20 instructions for each task. The prompts generated by the instructions are available in our repository. We provide more details on our data preparation in Section \textit{\textcolor{VulBinCode}{\textbf{DeBinVul}} Dataset Preparation} in Appendix. \ref{9_appendix}

\label{qa4binvul}





\subsection{Step 4: Fine Tuning Process}


\paragraph{\textbf{Tokenization of Decompiled Code.}}

We use a byte-pair encoding (BPE) tokenizer, common in natural language and programming language processing, to efficiently manage large vocabularies by merging frequent byte or character pairs into tokens. This approach reduces vocabulary size while preserving common patterns, balancing granularity and efficiency for handling diverse language data. From each function \( f \), we extract a set of tokens \( T \), trimming the input size to 512 tokens. We also add special tokens \( <BOS> \) and \( <EOS> \) at the start and end of the program, respectively, and pad sequences shorter than 32000 tokens with \( <PAD> \). The tokenized decompiled code is then used as input for the model.


\paragraph{\textbf{Model Training and Optimization.}}
In this work, we explore the application of generative language models to four tasks: i) Vulnerability identification, ii) Vulnerability classification, iii) Function name prediction, and iv) Description generation. Although the first two tasks are typically classification tasks (binary and multiclass, respectively), we convert all four tasks into generative ones by leveraging our model's instruction-following capability. Specifically, the model outputs ``Yes/No'' for vulnerability identification, generates a ``CWE-XXX'' code for classification, predicts a single token for the function name, and produces multiple tokens for description generation, enabling a unified multitask approach.

    \label{sec:approach}

    \section{Evaluation}
    

In this section, we evaluate the effectiveness of our proposed dataset \textcolor{VulBinCode}{\textbf{DeBinVul}} by benchmarking them on state-of-the-art LLMs and comparing their performance on the test set before and after fine-tuning. We evaluate our proposed system  to answer the following Research Questions (RQs):

\textbf{RQ1:} Using our instruction-based dataset \textcolor{VulBinCode}{\textbf{DeBinVul}}, how effectively can it be used to identify and classify binaries using different LLMs?


\textbf{RQ2:} How do the trained models with our dataset perform in function name prediction and description?

\color{black}
\textbf{RQ3:} Are the current LLMs generalized enough to analyze vulnerabilities in different architectures and optimization beyond their presence in their dataset?
\color{black}


\subsection{Evaluation Metrics}
\label{sec:model_metrics}


Our evaluation uses various task-specific metrics. For example, we used accuracy, precision, recall, and F1 scores for vulnerability identification and detection tasks in decompiled code. Acc.V refers to accuracy when all the input functions are vulnerable, and Acc.B refers to accuracy when all the input functions to the model are benign or non-vulnerable. However, we rely on metrics like BLEU (B.), Rouge-L (R.L), BERTScore (B.Score), and Semantic Similarity (Sim.)for function name prediction and description generation tasks. We put more details of the evaluation metrics in the Section Evaluation Metrics in the Appendix. \ref{9_appendix}


\subsection{Experimental Analysis}

\textbf{Experimental Setup.}
For our evaluations, we split our \textcolor{VulBinCode}{\textbf{DeBinVul}} dataset into 80\% training, 10\% validation, and 10\% testing. The training data included source code from the NVD dataset up to December 2021 to ensure that test data always followed the training data chronologically. We trained all benchmark models on an NVIDIA DGX server with an AMD EPYC 7742 64-Core processor, 1TB of RAM, and 8 NVIDIA A100 GPUs. The model was trained for four epochs with a maximum token length of 512, a learning rate of $2e^{-5}$, and a batch size of 4 for our 7B parameter model. A beam size of 1 and a temperature value of 1.0 were used for the generation task.




\begin{table}[t]
\centering
\caption{RQ1: Vulnerability identification task comparison between \textit{state-of-the-art} LLMs \textit{vs.} those trained on our dataset, \textcolor{VulBinCode}{\textbf{DeBinVul}}, referred as \textcolor{VulBinCode}{\textbf{DBVul}} in table.}
\begin{tabular}{p{0.09\textwidth}
                p{0.06\textwidth}
                p{0.02\textwidth}
                p{0.02\textwidth}
                p{0.02\textwidth}
                p{0.02\textwidth}
                p{0.03\textwidth}
                p{0.03\textwidth}}

\toprule \toprule
\textbf{Model}             & \textbf{Training} & \textbf{Acc} & \textbf{Pre.} & \textbf{Rec.} & \textbf{F1} & \textbf{Acc.V} & \textbf{Acc.B} \\ \midrule \midrule
\multirow{2}{*}{\textbf{CodeLLaMa}} & -              & 0.56         & 0.6           & 0.78          & 0.68        & 0.78           & 0.23           \\
                           & \textcolor{VulBinCode}{\textbf{DBVul}}       & \textcolor{DarkGreen}{\textbf{0.85}}         & \textcolor{DarkGreen}{\textbf{0.89}}          & \textcolor{DarkGreen}{\textbf{0.86}}          & \textcolor{DarkGreen}{\textbf{0.87}}        & \textcolor{DarkGreen}{\textbf{0.86}}           & \textcolor{DarkGreen}{\textbf{0.84}}           \\ \midrule
\multirow{2}{*}{\textbf{CodeGen2}}  & -              & 0.59         & 0.65          & 0.83          & 0.73        & 0.83           & 0.13           \\
                           & \textcolor{VulBinCode}{\textbf{DBVul}}        & \textcolor{DarkGreen}{\textbf{0.91}}         & \textcolor{DarkGreen}{\textbf{0.93}}          & \textcolor{DarkGreen}{\textbf{0.94}}          & \textcolor{DarkGreen}{\textbf{0.94}}        & \textcolor{DarkGreen}{\textbf{0.94}}           & \textcolor{DarkGreen}{\textbf{0.86}}           \\ \midrule
\multirow{2}{*}{\textbf{Mistral}}   & -              & 0.48         & 0.71          & 0.42          & 0.53        & 0.42           & 0.61           \\
                           & \textcolor{VulBinCode}{\textbf{DBVul}}        & \textcolor{DarkGreen}{\textbf{0.89}}         & \textcolor{DarkGreen}{\textbf{0.95}}          & \textcolor{DarkGreen}{\textbf{0.88}}          & \textcolor{DarkGreen}{\textbf{0.91}}        & \textcolor{DarkGreen}{\textbf{0.88}}           & \textcolor{DarkGreen}{\textbf{0.9}}            \\ \midrule
\multirow{2}{*}{\textbf{StarCoder}} & -              & 0.59         & 0.6           & 0.97          & 0.74        & 0.97           & 0.01           \\
                           & \textcolor{VulBinCode}{\textbf{DBVul}}        & \textcolor{DarkGreen}{\textbf{0.89}}         & \textcolor{DarkGreen}{\textbf{0.91}}          & 0.93          & \textcolor{DarkGreen}{\textbf{0.92}}        & 0.93           & \textcolor{DarkGreen}{\textbf{0.80}}           \\ \midrule
\multirow{2}{*}{\textbf{LLaMa 3}}   & -              & 0.57         & 0.7           & 0.68          & 0.69        & 0.68           & 0.34           \\
                           & \textcolor{VulBinCode}{\textbf{DBVul}}        & \textcolor{DarkGreen}{\textbf{0.91}}         & \textcolor{DarkGreen}{\textbf{0.94}}          & \textcolor{DarkGreen}{\textbf{0.93}}          & \textcolor{DarkGreen}{\textbf{0.93}}        & \textcolor{DarkGreen}{\textbf{0.93}}           & \textcolor{DarkGreen}{\textbf{0.87}}           \\ \bottomrule
\end{tabular}
\label{tab:base_ident}
\end{table}



\begin{table}[b]
\centering
\caption{RQ1: Vulnerability classification task comparison between base LLMs \textit{vs.} those trained on our dataset, \textcolor{VulBinCode}{\textbf{DeBinVul}}, referred as \textcolor{VulBinCode}{\textbf{DBVul}} in table.}
\begin{tabular}{p{0.05\textwidth}
                p{0.07\textwidth}
                p{0.07\textwidth}
                p{0.04\textwidth}
                p{0.06\textwidth}
                p{0.06\textwidth}}
\toprule \toprule
                 & \textbf{C.LLaMa} & \textbf{CodeGen2} & \textbf{Mistral} & \textbf{LLaMa3} & \textbf{St.Coder} \\ \midrule \midrule
Base    & 0.04               & 0                 & 0.04             & 0.02             & 0.03               \\
\textcolor{VulBinCode}{\textbf{DBVul}} & \textbf{0.81}\textcolor{DarkGreen}{$\uparrow$}      & \textbf{0.85}\textcolor{DarkGreen}{$\uparrow$}     & \textbf{0.83}\textcolor{DarkGreen}{$\uparrow$}    & \textbf{0.9}\textcolor{DarkGreen}{$\uparrow$}     & \textbf{0.84}\textcolor{DarkGreen}{$\uparrow$}      \\ \bottomrule
\end{tabular}
\label{tab:classification_f1}
\end{table}

\subsection{Answering Research Question 1} 
In answering RQ1, we investigate the effectiveness of the impact of the proposed dataset in analyzing binary code for four tasks, namely i) Vulnerability Identification, ii) Vulnerability Classification, iii) Function Name Prediction, and iv) Description of Code Objective. Throughout answering all our RQs for vulnerability identification and classification, we use Accuracy, Precision, Recall, and F1 scores. We use BLEU, Rouge-L, BERTScore, and Cosine Similarity for function name prediction and description generation. To answer RQ1, we only used O0 optimization on x86 architecture. Table \ref{tab:base_ident} shows the baseline comparison of the identification task on binary code. The \textit{Training} column with value Base implies the results were before training the model, and Our DS denotes after the LLM was fine-tuned with our dataset. Overall all the LLMs, when trained with our proposed dataset, show an improvement of F1 score of 18\% or higher. While we see that without training, CodeGen2 and StarCoder outperform by 59\% in identifying vulnerability. However, since this is a binary task, it is very close to a randomized guess, which is approximately 50\%. Moreover, if we see the individual accuracy only on vulnerable and only on non-vulnerable or benign code, we can see that some models like CodeGen2 \cite{nijkamp2023codegen2}, StarCoder \cite{li2023starcoder}, and CodeLLaMa \cite{roziere2023code} have significantly lower accuracy (Starcoder: 70\% lower) in identifying the non-vulnerable or benign functions while maintaining a higher accuracy in identifying the vulnerable functions. This phenomenon concludes that these models prefer to determine that most functions are vulnerable, hence the identification imbalance. However, after all the models were individually trained on our proposed dataset, we see an overall increase in the accuracy and F1 score, and CodeGen2 and LLaMa 3 top on this task with an accuracy of 91\%, which is almost a 30\% improvement from the baseline models. Furthermore, when we see the accuracy on only vulnerable and only benign functions, we see that, for both of the cases, the performance has remained high where CodeGen is 94\% successful at finding the vulnerable functions and LLaMa 3 is 87\% successful in finding the non-vulnerable or the benign functions. For classification, in Table \ref{tab:classification_f1}, we show the F1 score comparison. We can see that all the base models have a classification F1 score of less that 5\%, and interestingly, while CodeGen2 is a code-based LLM, it shows a score of 0 (zero) for vulnerability classification. Table \ref{tab:classification_details} compares all the CWEs in different models more in-depth. We provide more details on the classification task in Subsection \textcolor{black}{\textit{Further Discussion on RQ1} in Appendix}.

\subsection{Answering Research Question 2}
\begin{figure}
    \centering
    \includegraphics[width=\linewidth]{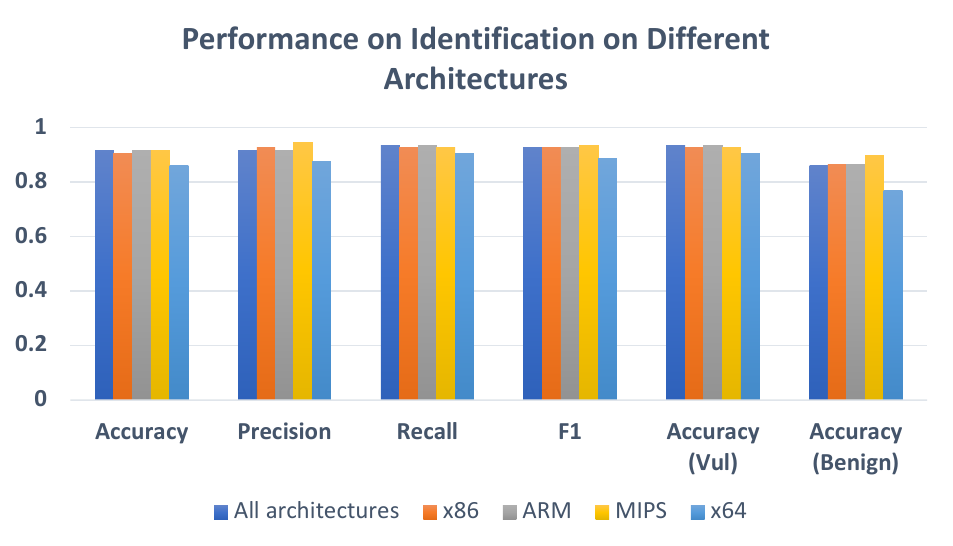}
    \caption{RQ2: Comparison of identification task on different architectures on decompiled binaries}
    \label{fig:ident_all_arch}
\end{figure}

\begin{figure}
    \centering
    \includegraphics[width=1\linewidth]{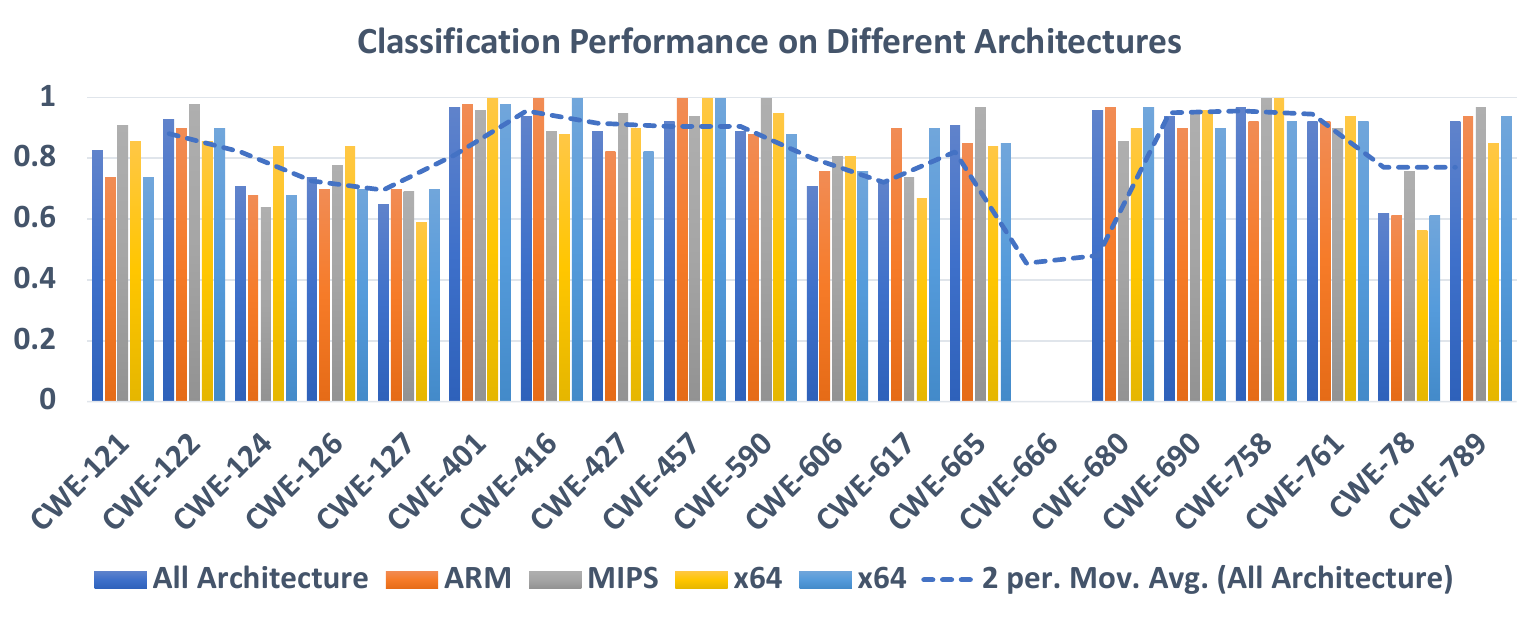}
    \caption{RQ2: Comparison of Classification on the combination of all architectures and individual architectures}
    \label{fig:class_all_arch}
\end{figure}

Our aim in answering RQ2 is to analyze one of the top-performing models to understand the performance of different architectures. Hence, we selected CodeLLaMa for this task to analyze the vulnerability of decompiled code. Here, we again train the based models on the same four tasks we performed in RQ1. However, RQ2 differs from RQ1, using a multi-architecture compilation of source code into decompiled code. For identification in Figure \ref{fig:ident_all_arch}, we see that the performance is close to approximately 90\% when we test by combining all the architectures. However, we see an improvement in Precision, F1, and accuracy in non-vulnerable or benign functions for MIPS. Moreover, we see a significant performance drop of 2-3\% overall metrics for x64 architecture, wherein the performance of x86, ARM, and MIPS remains relatively similar. Similarly, we see mixed results on F1 score for multiclass classification of CWEs in Figure \ref{fig:class_all_arch}. For example, on CWe-121, CWE-122, CWE-427, CWE-665, CWE-758, CWE-758 and, CWE-789 MIPS performs the higher. However, for CWE-401, CWE-690, and CWE-761, we see a relatively stable performance across all architectures. An interesting observation from Figure \ref{fig:class_all_arch} is that, for CWE-666, the F1 score goes down to zero, which implies a limitation of our dataset on CWE-666. If we follow the trend line of the moving average for "All Architecture" we see that, overall, the model performs lower for CWE-126, CWE-617, CWE-666, and CWE-78 while maintaining good performance on the other CWEs.

For the task of function name prediction and description generation in Table \ref{tab:all_name_desc}, the Cosine Similarity score shows a lower performance of x64 of 78\% while MIPS and the combination of all architectures shows a 4\% improvement of 82\% on this task. For the Description generation of decompiled code, we see a more stable score, where ARM, MIPS, and x64 top on a 76\% similarity score, wherein x86 shows a merely 2\% lower performance of 74\%.

\begin{table}[t]
\centering
\caption{RQ1: Performance of LLMs on Function Name Prediction and Description Generation tasks. }
\begin{tabular}{p{0.02\textwidth}
                p{0.06\textwidth}
                p{0.09\textwidth}
                p{0.02\textwidth}
                p{0.02\textwidth}
                p{0.02\textwidth}
                p{0.02\textwidth}
                p{0.02\textwidth}}
\toprule
\textbf{Task}                              & \textbf{Train} & \textbf{Model}             & \textbf{B.} & \textbf{R.L.} & \multicolumn{2}{c}{\textbf{B.Score}} & \textbf{Sim} \\ \midrule
\multirow{11}{*}{\rotatebox{90}{Function Name Prediction}} &                &                            &             &               & Prec.             & F1               &              \\
                                           & Our DS         & \multirow{2}{*}{CodeLLaMa} & 0.65        & 0.75          & 0.96              & 0.96             & 0.81         \\
                                           & Base           &                            & 0.00        & 0.00          & 0.77              & 0.78             & 0.10         \\ \cmidrule(l){2-8} 
                                           & Our DS         & \multirow{2}{*}{Mistral}   & 0.62        & 0.69          & 0.95              & 0.95             & 0.76         \\
                                           & Base           &                            & 0.00        & 0.01          & 0.78              & 0.79             & 0.40         \\ \cmidrule(l){2-8} 
                                           & Our DS         & \multirow{2}{*}{CodeGen2}  & 0.64        & 0.72          & 0.96              & 0.96             & 0.79         \\
                                           & Base           &                            & 0.02        & 0.01          & 0.78              & 0.79             & 0.15         \\ \cmidrule(l){2-8} 
                                           & Our DS         & \multirow{2}{*}{StarCoder} & 0.53        & 0.69          & 0.94              & 0.95             & 0.79         \\
                                           & Base           &                            & 0.00        & 0.00          & 0.78              & 0.79             & 0.40         \\ \cmidrule(l){2-8} 
                                           & Our DS         & \multirow{2}{*}{LLaMa 3}   & 0.66        & 0.77          & 0.97              & 0.97             & 0.83         \\
                                           & Base           &                            & 0.01        & 0.01          & 0.78              & 0.79             & 0.34         \\ \midrule
\multirow{10}{*}{\rotatebox{90}{Description}}              & Our DS         & \multirow{2}{*}{CodeLLaMa} & 0.12        & 0.29          & 0.89              & 0.88             & 0.77         \\
                                           & Base           &                            & 0.01        & 0.17          & 0.78              & 0.79             & 0.39         \\ \cmidrule(l){2-8} 
                                           & Our DS         & \multirow{2}{*}{Mistral}   & 0.13        & 0.30          & 0.89              & 0.88             & 0.78         \\
                                           & Base           &                            & 0.03        & 0.18          & 0.80              & 0.81             & 0.48         \\ \cmidrule(l){2-8} 
                                           & Our DS         & \multirow{2}{*}{CodeGen2}  & 0.11        & 0.28          & 0.88              & 0.88             & 0.77         \\
                                           & Base           &                            & 0.00        & 0.02          & 0.76              & 0.77             & 0.18         \\ \cmidrule(l){2-8} 
                                           & Our DS         & \multirow{2}{*}{StarCoder} & 0.09        & 0.25          & 0.83              & 0.85             & 0.71         \\
                                           & Base           &                            & 0.00        & 0.02          & 0.76              & 0.77             & 0.18         \\ \cmidrule(l){2-8} 
                                           & Our DS         & \multirow{2}{*}{LLaMa 3}   & 0.13        & 0.30          & 0.89              & 0.88             & 0.78         \\
                                           & Base           &                            & 0.02        & 0.18          & 0.83              & 0.83             & 0.50         \\ \bottomrule
\end{tabular}
\label{tab:base_name_desc}
\end{table}

\subsection{Answering Research Question 3}


Our goal in answering RQ3 is to demonstrate how well our CodeLLaMa performs when trained on a subset of architectures and tested on a different subset of architectures for function name prediction and description generation tasks. In Table \ref{tab:gen_arch}, the column ``Train'' depicts the set of architectures that were present during the training and the ``Tes'' column defines the set of architectures that were present during the testing. however, we kept some overlap in the architectures between the training and testing for comparison tasks. $All - x64$ defines that the model trained will all three architectures except $x64$, and $ARM + x86$ defines that the model was only trained on ARM and x86 architecture. For function name prediction, on Table \ref{tab:gen_arch}, we can see that when the model was trained without x64, we see a very slight performance drop of only 1\% on the Cosine Similarity score when tested on x64. However, when the model was trained on ARM and x86, we see that for x86, there was a 4\% drop in the performance compared to ARM, while x86 was still in the training data. Furthermore, for description, when the model was trained with $All - x64$, the performance of x64 only dropped by 2\% for the Cosine Similarity score, and when the model was trained on $ARM + x86$, and tested with ``Al'' we see almost no performance change. Furthermore, we also tested generalizability on O0 and O3 optimization levels on function name prediction and description tasks. For both tasks, the model was trained on O0 optimization and tested on the 03 optimization level. We see a mere 1\% improvement when the model was trained and tested on the same optimization level. From this analysis, we can safely conclude that using different architectures has almost little to no effect on function name prediction and description generation tasks.




\begin{table}[b]
\centering
\caption{RQ2: Function Name Prediction and Description when source code compiled on different architectures.}
\begin{tabular}{@{}llcccc@{}}
\toprule
\textbf{Task}                                                        & \textbf{Arch.}      & \multicolumn{1}{l}{\textbf{BLEU}} & \multicolumn{1}{l}{\textbf{Rouge-L}} & \multicolumn{1}{l}{\textbf{BERTScore}} & \multicolumn{1}{l}{\textbf{Sim.}} \\ \midrule
\multicolumn{1}{l|}{\multirow{5}{*}{\textbf{\rotatebox{90}{Func. Name}}}} & \textbf{All} & 0.64                          & 0.75                           & 0.96                             & 0.82                                         \\
\multicolumn{1}{l|}{}                                                & \textbf{x86}               & 0.62                          & 0.72                           & 0.96                             & 0.80                                         \\
\multicolumn{1}{l|}{}                                                & \textbf{ARM}               & 0.67                          & 0.75                           & 0.96                             & 0.81                                         \\
\multicolumn{1}{l|}{}                                                & \textbf{MIPS}              & 0.68                          & 0.77                           & 0.96                             & 0.82                                         \\
\multicolumn{1}{l|}{}                                                & \textbf{x64}               & 0.61                          & 0.71                           & 0.95                             & 0.78                                          \\ \midrule
\multicolumn{1}{l|}{\multirow{5}{*}{\textbf{\rotatebox{90}{Description}}}}           & \textbf{All} & 0.12                        & 0.30                          & 0.88                             & 0.75                                         \\
\multicolumn{1}{l|}{}                                                & \textbf{x86}               & 0.11                        & 0.29                         & 0.88                             & 0.74                                         \\
\multicolumn{1}{l|}{}                                                & \textbf{ARM}               & 0.11                        & 0.30                         & 0.88                             & 0.76                                         \\
\multicolumn{1}{l|}{}                                                & \textbf{MIPS}              & 0.12                        & 0.31                          & 0.88                             & 0.76                                         \\
\multicolumn{1}{l|}{}                                                & \textbf{x64}               & 0.13                        & 0.30                         & 0.88                              & 0.76                                         \\ \bottomrule 
\end{tabular}
\label{tab:all_name_desc}
\end{table}

\begin{table}[t]
\centering
\caption{RQ3: Generalizability Testing with Different Architectures and Optimization Levels on predicting function name and description generation.}
\begin{tabular}{p{0.02\textwidth}
                p{0.03\textwidth}
                p{0.03\textwidth}
                p{0.03\textwidth}
                p{0.03\textwidth}
                p{0.03\textwidth}
                p{0.01\textwidth}
                p{0.01\textwidth}}
\hline
\textbf{Task}                                                                               & \textbf{Train}                                                                             & \textbf{Test} & \textbf{B.}             & \textbf{R.L.}          & \multicolumn{2}{c}{\textbf{B.Score}} & \textbf{Sim}            \\ \hline
                                                                                            &                                                                                            &               &                          &                          & Pre.     & F1                          &                          \\ \hline
\multicolumn{1}{l|}{\multirow{7}{*}{\begin{tabular}[c]{@{}l@{}}\rotatebox{90}{Function Name}\end{tabular}}} & \multicolumn{1}{l|}{\multirow{2}{*}{\begin{tabular}[c]{@{}l@{}}All\\ -x64\end{tabular}}}   & ARM           & 0.61                     & 0.72                     & 0.96     & 0.96                        & 0.79                     \\ \cline{3-8} 
\multicolumn{1}{l|}{}                                                                       & \multicolumn{1}{l|}{}                                                                      & x64           & 0.61                     & 0.71                     & 0.96     & 0.96                        & 0.78                     \\ \cline{2-8} 
\multicolumn{1}{l|}{}                                                                       & \multicolumn{1}{l|}{\multirow{3}{*}{\begin{tabular}[c]{@{}l@{}}ARM \\ + x86\end{tabular}}} & All           & 0.62                     & 0.70                     & 0.96     & 0.96                        & 0.77                     \\ \cline{3-8} 
\multicolumn{1}{l|}{}                                                                       & \multicolumn{1}{l|}{}                                                                      & x86           & 0.60                     & 0.67                     & 0.95     & 0.95                        & 0.75                     \\ \cline{3-8} 
\multicolumn{1}{l|}{}                                                                       & \multicolumn{1}{l|}{}                                                                      & ARM           & 0.64                     & 0.72                     & 0.96     & 0.96                        & 0.79                     \\ \cline{2-8} 
\multicolumn{1}{l|}{}                                                                       & \multicolumn{1}{l|}{\multirow{2}{*}{O0}}                                                   & O0            & 0.30                     & 0.36                     & 0,89     & 0.89                        & 0.44                     \\ \cline{3-8} 
\multicolumn{1}{l|}{}                                                                       & \multicolumn{1}{l|}{}                                                                      & O3            & \multicolumn{1}{l}{0.28} & \multicolumn{1}{l}{0.35} & 0.89     & \multicolumn{1}{c}{0.88}    & \multicolumn{1}{c}{0.44} \\ \hline
\multicolumn{1}{l|}{\multirow{6}{*}{\rotatebox{90}{Description}}}                                                 & \multicolumn{1}{l|}{\multirow{2}{*}{\begin{tabular}[c]{@{}l@{}}All\\ -x64\end{tabular}}}   & ARM           & 0.12                     & 0.31                     & 0.89     & 0.89                        & 0.77                     \\ \cline{3-8} 
\multicolumn{1}{l|}{}                                                                       & \multicolumn{1}{l|}{}                                                                      & x64           & 0.12                     & 0.29                     & 0.88     & 0.88                        & 0.75                     \\ \cline{2-8} 
\multicolumn{1}{l|}{}                                                                       & \multicolumn{1}{l|}{\multirow{2}{*}{\begin{tabular}[c]{@{}l@{}}ARM \\ + x86\end{tabular}}} & All           & 0.11                     & 0.29                     & 0.89     & 0.88                        & 0.75                     \\ \cline{3-8} 
\multicolumn{1}{l|}{}                                                                       & \multicolumn{1}{l|}{}                                                                      & x86           & 0.10                     & 0.29                     & 0.89     & 0.88                        & 0.75                     \\ \cline{2-8} 
\multicolumn{1}{l|}{}                                                                       & \multicolumn{1}{l|}{\multirow{2}{*}{O0}}                                                   & O0            & 0.10                     & 0.25                     & 0.89     & 0.89                        & 0.70                     \\ \cline{3-8} 
\multicolumn{1}{l|}{}                                                                       & \multicolumn{1}{l|}{}                                                                      & O3            & \multicolumn{1}{c}{0.08} & \multicolumn{1}{c}{0.25} & 0.87     & \multicolumn{1}{c}{0.87}    & \multicolumn{1}{c}{0.69} \\ \hline
\end{tabular}
\label{tab:gen_arch}
\end{table}

To evaluate the generalizability of Large Language Models (LLMs) in real-world scenarios, we conducted a small-scale experiment using a generalized instruction-based dataset. Specifically, we tested Mistral and LLaMA 3 on the Stanford Alpaca dataset \cite{alpaca}, performing inference on the base models prior to training with our dataset. Initial cosine similarity scores were 0.67 for Mistral and 0.73 for LLaMA 3. After training the models on our proposed dataset, we reassessed performance on the Stanford Alpaca dataset, observing that cosine similarity scores for Mistral and LLaMA 3 dropped to 0.56 and 0.70, respectively. The notable decrease in Mistral's performance is likely due to its smaller model size (2B parameters), which led to catastrophic forgetting when trained on new data, whereas the 7B-parameter LLaMA 3 retained much of its prior learning. Additionally, we conducted an N-day vulnerability analysis, where LLaMA 3 and Mistral identified 15 and 6 N-day vulnerabilities, respectively.

    \section{Related Work}
    
Recent advances in binary vulnerability detection have focused on leveraging intermediate representations and deep learning techniques to address the challenges posed by code reuse. VulHawk \cite{luo2023vulhawk} employed an intermediate representation-based approach using RoBERTa \cite{liu2019roberta} to embed binary code and applied a progressive search strategy for identifying vulnerabilities in similar binaries. Asteria-Pro \cite{yang2023asteria} utilized LSTM \cite{hochreiter1997long} for large-scale binary similarity detection, while VulANalyzeR \cite{li2023vulanalyzer} proposed an attention-based method with Graph Convolution \cite{kipf2017semi} and Control Flow Graphs to classify vulnerabilities and identify root causes. QueryX \cite{han2023queryx} took a different approach by converting binaries into static source code through symbolic analysis and decompilation to detect bugs in commercial Windows kernels. In the realm of code summarization for decompiled binaries, Kawsan et al. \cite{al2023extending} fine-tuned the CodeT5 model \cite{wang2021codet5} on decompiled function-summary pairs, while HexT5 \cite{xiong2023hext5} extended CodeT5 for tasks like code summarization and variable recovery. BinSum \cite{jin2023binary} introduced a binary code summarization dataset and evaluated LLMs such as GPT-4 \cite{OpenAI2023GPT4}, Llama-2 \cite{touvron2023llama}, and Code-LlaMa \cite{roziere2023code} across various optimization levels and architectures. Additionally, Asm2Seq \cite{taviss2024asm2seq} focused on generating textual summaries of assembly functions for vulnerability analysis, specifically targeting x86 assembly instructions.

    \section{Conclusion}

In this study, we present a comprehensive investigation of large language models (LLMs) for the classification and identification of vulnerabilities in decompiled code and source code to determine the semantic gap. The primary contribution of our work is the development of the \textcolor{VulBinCode}{\textbf{DeBinVul}} dataset, an extensive instruction-based resource tailored for vulnerability identification, classification, function name prediction, and description across four architectures and two optimization levels. Our experiments demonstrate that \textcolor{VulBinCode}{\textbf{DeBinVul}} significantly improves vulnerability identification and classification by up to 30\% compared to baseline models on the x86 architecture. Additionally, we provide an in-depth analysis of how different LLMs perform across various computer architectures for all four tasks. We also evaluated how our proposed dataset aids LLMs in generalizing across different architectures and optimization levels. Our results indicate that the LLMs maintained consistent performance even when exposed to new architectures or optimization methods not included in the training data.

    \bibliography{aaai25}

\begin{thebibliography}{37}
\providecommand{\natexlab}[1]{#1}

\bibitem[{Akamai(2023)}]{akamai}
Akamai. 2023.
\newblock XZ Utils Backdoor — Everything You Need to Know, and What You Can Do.
\newblock Accessed: 2024-05-19.

\bibitem[{Al-Kaswan et~al.(2023)Al-Kaswan, Ahmed, Izadi, Sawant, Devanbu, and van Deursen}]{al2023extending}
Al-Kaswan, A.; Ahmed, T.; Izadi, M.; Sawant, A.~A.; Devanbu, P.; and van Deursen, A. 2023.
\newblock Extending source code pre-trained language models to summarise decompiled binarie.
\newblock In \emph{2023 IEEE International Conference on Software Analysis, Evolution and Reengineering (SANER)}, 260--271. IEEE.

\bibitem[{Balakrishnan and Reps(2010)}]{balakrishnan2010wysinwyx}
Balakrishnan, G.; and Reps, T. 2010.
\newblock Wysinwyx: What you see is not what you execute.
\newblock \emph{ACM Transactions on Programming Languages and Systems (TOPLAS)}, 32(6): 1--84.

\bibitem[{Brunsfeld et~al.(2018)Brunsfeld, Thomson, Vera, Hlynskyi, Turnbull, Clem, and Muller}]{brunsfeld2018tree}
Brunsfeld, M.; Thomson, P.; Vera, J.; Hlynskyi, A.; Turnbull, P.; Clem, T.; and Muller, A. 2018.
\newblock tree-sitter/tree-sitter: v0. 20.0.

\bibitem[{Burk et~al.(2022)Burk, Pagani, Kruegel, and Vigna}]{burk2022decomperson}
Burk, K.; Pagani, F.; Kruegel, C.; and Vigna, G. 2022.
\newblock Decomperson: How humans decompile and what we can learn from it.
\newblock In \emph{31st USENIX Security Symposium (USENIX Security 22)}, 2765--2782.

\bibitem[{Eschweiler et~al.(2016)Eschweiler, Yakdan, Gerhards-Padilla et~al.}]{eschweiler2016discovre}
Eschweiler, S.; Yakdan, K.; Gerhards-Padilla, E.; et~al. 2016.
\newblock Discovre: Efficient cross-architecture identification of bugs in binary code.
\newblock In \emph{Ndss}, volume~52, 58--79.

\bibitem[{Google(2023)}]{Google2023Gemini}
Google. 2023.
\newblock Gemini.
\newblock Accessed: 2024-05-19.

\bibitem[{Han et~al.(2023)Han, Kyea, Jin, Kang, Pak, and Yun}]{han2023queryx}
Han, H.; Kyea, J.; Jin, Y.; Kang, J.; Pak, B.; and Yun, I. 2023.
\newblock QueryX: Symbolic Query on Decompiled Code for Finding Bugs in COTS Binaries.
\newblock In \emph{2023 IEEE Symposium on Security and Privacy (SP)}, 3279--312795. IEEE.

\bibitem[{Hochreiter and Schmidhuber(1997)}]{hochreiter1997long}
Hochreiter, S.; and Schmidhuber, J. 1997.
\newblock Long short-term memory.
\newblock \emph{Neural computation}, 9(8): 1735--1780.

\bibitem[{Hu, Liang, and Chen(2024)}]{hu2024degpt}
Hu, P.; Liang, R.; and Chen, K. 2024.
\newblock DeGPT: Optimizing Decompiler Output with LLM.
\newblock In \emph{Proceedings 2024 Network and Distributed System Security Symposium (2024). https://api. semanticscholar. org/CorpusID}, volume 267622140.

\bibitem[{Huang and Leng(1999)}]{huang1999generalized}
Huang, J.-C.; and Leng, T. 1999.
\newblock Generalized loop-unrolling: a method for program speedup.
\newblock In \emph{Proceedings 1999 IEEE Symposium on Application-Specific Systems and Software Engineering and Technology. ASSET'99 (Cat. No. PR00122)}, 244--248. IEEE.

\bibitem[{Jiang et~al.(2023)Jiang, Sablayrolles, Mensch, Bamford, Chaplot, de~las Casas, Bressand, Lengyel, Lample, Saulnier et~al.}]{jiang2023mistral}
Jiang, A.; Sablayrolles, A.; Mensch, A.; Bamford, C.; Chaplot, D.; de~las Casas, D.; Bressand, F.; Lengyel, G.; Lample, G.; Saulnier, L.; et~al. 2023.
\newblock Mistral 7B (2023).
\newblock \emph{arXiv preprint arXiv:2310.06825}.

\bibitem[{Jin et~al.(2023)Jin, Larson, Yang, and Lin}]{jin2023binary}
Jin, X.; Larson, J.; Yang, W.; and Lin, Z. 2023.
\newblock Binary code summarization: Benchmarking chatgpt/gpt-4 and other large language models.
\newblock \emph{arXiv preprint arXiv:2312.09601}.

\bibitem[{Kipf and Welling(2017)}]{kipf2017semi}
Kipf, T.~N.; and Welling, M. 2017.
\newblock Semi-Supervised Classification with Graph Convolutional Networks.
\newblock In \emph{International Conference on Learning Representations (ICLR)}.

\bibitem[{Lee et~al.(2015)Lee, Song, Jang, Wang, Kim, Lu, and Lee}]{lee2015preventing}
Lee, B.; Song, C.; Jang, Y.; Wang, T.; Kim, T.; Lu, L.; and Lee, W. 2015.
\newblock Preventing Use-after-free with Dangling Pointers Nullification.
\newblock In \emph{NDSS}.

\bibitem[{Li et~al.(2023{\natexlab{a}})Li, Ding, Tian, Fung, Charland, Ou, Song, and Chen}]{li2023vulanalyzer}
Li, L.; Ding, S.~H.; Tian, Y.; Fung, B.~C.; Charland, P.; Ou, W.; Song, L.; and Chen, C. 2023{\natexlab{a}}.
\newblock VulANalyzeR: Explainable binary vulnerability detection with multi-task learning and attentional graph convolution.
\newblock \emph{ACM Transactions on Privacy and Security}, 26(3): 1--25.

\bibitem[{Li et~al.(2023{\natexlab{b}})Li, Allal, Zi, Muennighoff, Kocetkov, Mou, Marone, Akiki, Li, Chim et~al.}]{li2023starcoder}
Li, R.; Allal, L.~B.; Zi, Y.; Muennighoff, N.; Kocetkov, D.; Mou, C.; Marone, M.; Akiki, C.; Li, J.; Chim, J.; et~al. 2023{\natexlab{b}}.
\newblock StarCoder: may the source be with you!
\newblock \emph{arXiv preprint arXiv:2305.06161}.

\bibitem[{Lin(2004)}]{lin2004rouge}
Lin, C.-Y. 2004.
\newblock Rouge: A package for automatic evaluation of summaries.
\newblock In \emph{Text summarization branches out}, 74--81.

\bibitem[{Liu et~al.(2019)Liu, Ott, Goyal, Du, Joshi, Chen, Levy, Lewis, Zettlemoyer, and Stoyanov}]{liu2019roberta}
Liu, Y.; Ott, M.; Goyal, N.; Du, J.; Joshi, M.; Chen, D.; Levy, O.; Lewis, M.; Zettlemoyer, L.; and Stoyanov, V. 2019.
\newblock Roberta: A robustly optimized bert pretraining approach.
\newblock \emph{arXiv preprint arXiv:1907.11692}.

\bibitem[{Luo et~al.(2023)Luo, Wang, Wang, Tang, Xie, Zhou, Liu, and Lu}]{luo2023vulhawk}
Luo, Z.; Wang, P.; Wang, B.; Tang, Y.; Xie, W.; Zhou, X.; Liu, D.; and Lu, K. 2023.
\newblock VulHawk: Cross-architecture Vulnerability Detection with Entropy-based Binary Code Search.
\newblock In \emph{NDSS}.

\bibitem[{Mantovani et~al.(2022)Mantovani, Compagna, Shoshitaishvili, and Balzarotti}]{mantovani2022convergence}
Mantovani, A.; Compagna, L.; Shoshitaishvili, Y.; and Balzarotti, D. 2022.
\newblock The Convergence of Source Code and Binary Vulnerability Discovery--A Case Study.
\newblock In \emph{Proceedings of the 2022 ACM on Asia Conference on Computer and Communications Security}, 602--615.

\bibitem[{Nguyen et~al.(2020)Nguyen, Bardin, Bonichon, Groz, and Lemerre}]{nguyen2020binary}
Nguyen, M.-D.; Bardin, S.; Bonichon, R.; Groz, R.; and Lemerre, M. 2020.
\newblock Binary-level directed fuzzing for $\{$use-after-free$\}$ vulnerabilities.
\newblock In \emph{23rd International Symposium on Research in Attacks, Intrusions and Defenses (RAID 2020)}, 47--62.

\bibitem[{Nijkamp et~al.(2023)Nijkamp, Hayashi, Xiong, Savarese, and Zhou}]{nijkamp2023codegen2}
Nijkamp, E.; Hayashi, H.; Xiong, C.; Savarese, S.; and Zhou, Y. 2023.
\newblock Codegen2: Lessons for training llms on programming and natural languages.
\newblock \emph{arXiv preprint arXiv:2305.02309}.

\bibitem[{NSA(2019)}]{ghidra}
NSA. 2019.
\newblock Ghidra.
\newblock \url{https://ghidra-sre.org/}.
\newblock Software reverse engineering framework.

\bibitem[{OpenAI(2023)}]{OpenAI2023GPT4}
OpenAI. 2023.
\newblock GPT-4.
\newblock Accessed: 2024-05-19.

\bibitem[{Papineni et~al.(2002)Papineni, Roukos, Ward, and Zhu}]{papineni2002bleu}
Papineni, K.; Roukos, S.; Ward, T.; and Zhu, W.-J. 2002.
\newblock Bleu: a method for automatic evaluation of machine translation.
\newblock In \emph{Proceedings of the 40th annual meeting of the Association for Computational Linguistics}, 311--318.

\bibitem[{Qualys(2024)}]{qualys}
Qualys. 2024.
\newblock 2024 Midyear Threat Landscape Review.
\newblock \url{https://blog.qualys.com/vulnerabilities-threat-research/2024/08/06/2024-midyear-threat-landscape-review}.
\newblock Vulnerabilities Threat Research.

\bibitem[{Roziere et~al.(2023)Roziere, Gehring, Gloeckle, Sootla, Gat, Tan, Adi, Liu, Remez, Rapin et~al.}]{roziere2023code}
Roziere, B.; Gehring, J.; Gloeckle, F.; Sootla, S.; Gat, I.; Tan, X.~E.; Adi, Y.; Liu, J.; Remez, T.; Rapin, J.; et~al. 2023.
\newblock Code llama: Open foundation models for code.
\newblock \emph{arXiv preprint arXiv:2308.12950}.

\bibitem[{Sarkar(2000)}]{sarkar2000optimized}
Sarkar, V. 2000.
\newblock Optimized unrolling of nested loops.
\newblock In \emph{Proceedings of the 14th international conference on Supercomputing}, 153--166.

\bibitem[{Su et~al.(2024)Su, Xu, Huang, Zhang, and Zhang}]{su2024source}
Su, Z.; Xu, X.; Huang, Z.; Zhang, K.; and Zhang, X. 2024.
\newblock Source Code Foundation Models are Transferable Binary Analysis Knowledge Bases.
\newblock \emph{arXiv preprint arXiv:2405.19581}.

\bibitem[{Taori et~al.(2023)Taori, Gulrajani, Zhang, Dubois, Li, Guestrin, Liang, and Hashimoto}]{alpaca}
Taori, R.; Gulrajani, I.; Zhang, T.; Dubois, Y.; Li, X.; Guestrin, C.; Liang, P.; and Hashimoto, T.~B. 2023.
\newblock Stanford Alpaca: An Instruction-following LLaMA model.
\newblock \url{https://github.com/tatsu-lab/stanford_alpaca}.

\bibitem[{Taviss et~al.(2024)Taviss, Ding, Zulkernine, Charland, and Acharya}]{taviss2024asm2seq}
Taviss, S.; Ding, S.~H.; Zulkernine, M.; Charland, P.; and Acharya, S. 2024.
\newblock Asm2seq: Explainable assembly code functional summary generation for reverse engineering and vulnerability analysis.
\newblock \emph{Digital Threats: Research and Practice}, 5(1): 1--25.

\bibitem[{Touvron et~al.(2023)Touvron, Martin, Stone, Albert, Almahairi, Babaei, Bashlykov, Batra, Bhargava, Bhosale et~al.}]{touvron2023llama}
Touvron, H.; Martin, L.; Stone, K.; Albert, P.; Almahairi, A.; Babaei, Y.; Bashlykov, N.; Batra, S.; Bhargava, P.; Bhosale, S.; et~al. 2023.
\newblock Llama 2: Open foundation and fine-tuned chat models.
\newblock \emph{arXiv preprint arXiv:2307.09288}.

\bibitem[{Wang et~al.(2021)Wang, Wang, Joty, and Hoi}]{wang2021codet5}
Wang, Y.; Wang, W.; Joty, S.; and Hoi, S.~C. 2021.
\newblock {C}ode{T}5: Identifier-aware Unified Pre-trained Encoder-Decoder Models for Code Understanding and Generation.
\newblock In \emph{Proceedings of the 2021 Conference on Empirical Methods in Natural Language Processing}, 8696--8708. Online and Punta Cana, Dominican Republic: Association for Computational Linguistics.

\bibitem[{Xiong et~al.(2023)Xiong, Chen, Chen, Gao, Cheng, and Zhang}]{xiong2023hext5}
Xiong, J.; Chen, G.; Chen, K.; Gao, H.; Cheng, S.; and Zhang, W. 2023.
\newblock HexT5: Unified Pre-Training for Stripped Binary Code Information Inference.
\newblock In \emph{2023 38th IEEE/ACM International Conference on Automated Software Engineering (ASE)}, 774--786. IEEE.

\bibitem[{Yang et~al.(2023)Yang, Dong, Xiao, Cheng, Shi, Li, and Sun}]{yang2023asteria}
Yang, S.; Dong, C.; Xiao, Y.; Cheng, Y.; Shi, Z.; Li, Z.; and Sun, L. 2023.
\newblock Asteria-Pro: Enhancing Deep Learning-based Binary Code Similarity Detection by Incorporating Domain Knowledge.
\newblock \emph{ACM Transactions on Software Engineering and Methodology}, 33(1): 1--40.

\bibitem[{Zhang et~al.(2019)Zhang, Kishore, Wu, Weinberger, and Artzi}]{zhang2019bertscore}
Zhang, T.; Kishore, V.; Wu, F.; Weinberger, K.~Q.; and Artzi, Y. 2019.
\newblock Bertscore: Evaluating text generation with bert.
\newblock \emph{arXiv preprint arXiv:1904.09675}.

\end{thebibliography}
    
    \appendix
    
\section{Appendix}

\label{appen:injection}
In this appendix, we explain in detail the investigation injection process on different repositories and dataset collection methodologies.

\section{Source \& Decompiled Binary Code Vulnerability Semantic Gap: \textit{Investigating LLMs' Analytical Abilities}}

In this section, we are the first to empirically and pragmatically investigate the analytical abilities of \textit{state-of-the-art} LLMs in analyzing vulnerabilities in the decompiled binary code domain. To explore this, we randomly select \textit{200} pairs of vulnerable and non-vulnerable source code and decompiled binary code samples from our proposed dataset, \textcolor{VulBinCode}{\textbf{DeBinVul}} \footnote{Please refer to the Proposed Approach Section for details on the dataset.} for the task of classifying vulnerabilities. Specifically, we evaluate the ability of several LLMs, including \texttt{ChatGPT-4} \cite{OpenAI2023GPT4}, \texttt{Gemini} \cite{Google2023Gemini}, \texttt{CodeLLaMA-7B} \cite{roziere2023code}, Mistral \cite{jiang2023mistral}, LLaMa 3 \cite{touvron2023llama}, and CodeGen2 \cite{nijkamp2023codegen2}, to identify and classify vulnerabilities in both source code and decompiled binary code. Table \ref{tab:motivation} presents the comparison results and underscores the semantic vulnerability limitations of \textit{state-of-the-art} black-box LLMs in classifying \texttt{CWEs} in decompiled binary code, in contrast to source code, which presented moderately better results. The reported results in Table \ref{tab:motivation} focuses on \texttt{CWEs:} \texttt{787}, \texttt{416}, \texttt{20}, \texttt{125}, \texttt{476}, \texttt{190}, \texttt{119}, \texttt{798} and reports their average F1-scores.

A carefully designed prompt was used to leverage the generative capabilities of these LLMs, asking them to respond with a “Yes” or “No” regarding the presence of vulnerabilities. Additionally, the prompt required the LLMs to generate the corresponding CWE number if a vulnerability was detected. To classify the vulnerabilities, the prompt was adjusted to ensure that the LLMs only output the CWE category. The results from these LLMs are summarized and analyzed in Table \ref{tab:motivation}. The specific prompts used in this analysis are provided in Appendix \ref{9_appendix} \textit{\textcolor{black}{Prompts and Investigation}}.

\paragraph{Result Analysis}
The results from Table \ref{tab:motivation} in Appendix show that API-based models like GPT4 could accurately identify a vulnerability in decompiled binaries with an accuracy of 70\%, where Gemini is at 56\%. Moreover, open models like CodeLLaMa is at 61\%, Mistral is at 54\%, LLaMa 3 at 50\%, and CodeGen2 at 54\%.  Moreover, from Table \ref{tab:motivation}, we see that GPT-4 performs comparatively higher overall than other API-based or open-access models. To investigate the details of the identification task, we investigate how accurately these LLMs can classify each vulnerability category. The results show that all models were experts at identifying some vulnerabilities and failing at others. For example, GPT4, Gemini, Mistral, and LLaMa 3 produce higher performance on CWE-416, CodeLLaMa on CWE-20, LLaMa 3 on CWE-190, and CodeGen2 on CWe-476. However, one interesting observation from our analysis is that while CodeLLaMa, Mistral, and CodeGen 2 are moderately successful in identifying vulnerability, these LLMs fail to predict most CWEs. Therefore, we can conclude that CodeLLaMa has a very limited understanding of the vulnerability category. Furthermore, Table \ref{tab:motivation} provides more detailed numerical results, including Accuracy, Precision, Recall, and F1 score. Section \textit{Reasoning of Weak Performance in Investigation} in Appendix \ref{9_appendix} explains the reason behind the poor performance of LLMs when analyzing decompiled binaries.

\begin{figure*}[t]
    \centering
    \includegraphics[width=1\linewidth]{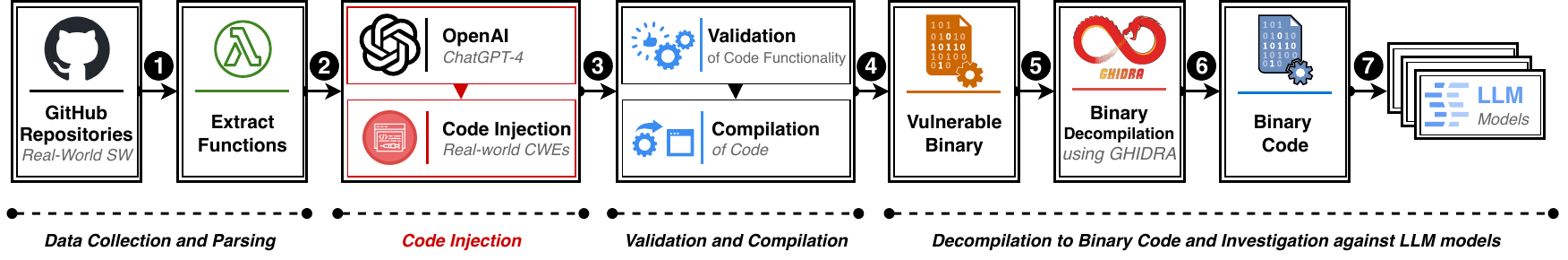}
    \caption{A high-level overview of our investigation process: From Vulnerability Injection to Vulnerability Analysis Using Decompiled Code}
    \label{fig:investigation}
\end{figure*}

\section{Vulnerability Injection Process}
This section shows the different prompts we used to investigate state-of-the-art LLMs and for instruction generation tasks. Table \ref{tab:insts} shows the three prompts we needed to generate the outcomes for vulnerability identification and classification tasks. However, we noticed some disparity when the models followed the same command. Initially, we created the prompt for GPT-4. However, when we used the same prompt for Gemini, we saw that it produced some extra outputs, which are the URLs of the CWEs. Therefore, we had to ground the behavior by updating and adding instructions with the prompt.

The other LLM Instruction columns refer to  CodeLLaMa, Mistral, CodeGen2, and LLaMa 3. When generating the output, the model was generating the CWE description. However, this time, we were unsuccessful in grounding that behavior using extra instruction. Therefore, when processing the output generated by the model, we wrote extra code to remove the description CodeLLaMa was generating.

\begin{table}[ht]
\centering
\caption{Number of functions and total counts in various top C/C++ general-purpose and IoT repositories where we injected 8 CWE vulnerabilities into some randomly selected functions.}
\begin{tabular}{lll}
\hline
\textbf{Repository Name} & \textbf{Domain}      & \textbf{Total} \\ \hline
\textbf{Linux Kernel}    & Operating System     & 3              \\
\textbf{Apache HTTPD}    & Web Server           & 7              \\
\textbf{OpenSSL}         & Security Library     & 2              \\
\textbf{FFmpeg}          & Multimedia Framework & 6              \\
\textbf{cURL}            & Data Transfer        & 5              \\
\textbf{MQTT}            & IoT Protocol         & 2              \\
\textbf{Zigbee}          & IoT Networking       & 8              \\
\textbf{Node.js}         & Runtime Environment  & 3              \\
\textbf{SQLite}          & Database             & 61             \\
\textbf{Json-C}          & Data Format          & 77             \\ \hline
\end{tabular}
\label{tab:repositories}

\end{table}


We initially extracted all the functions from these ten repositories to investigate their effectiveness. Then, we randomly selected some functions to inject vulnerabilities which are demonstrated in Table \ref{tab:motivation}. After injecting the vulnerabilities and fixing the compilation errors, we compile each repository into its binaries and decompile them back to the original code using Ghidra \cite{ghidra}. As a result, the decompiled versions of the functions for which we injected vulnerability are also vulnerable. Our analysis includes identification, classification, and function name prediction of the decompiled code. Table \ref{tab:repositories} summarizes the code we generated to investigate vulnerability across different LLMs.

\begin{table*}[t]
\centering
\caption{Performance of API-based and Open for Vulnerability Identification and Classification Tasks in Decompiled binaries.}
\begin{tabular}{@{}lllllllllllll@{}}
\toprule
\textbf{CWE Number}     & \multicolumn{4}{c}{\textbf{GPT-4}}              & \multicolumn{4}{c}{\textbf{Gemini}}            & \multicolumn{4}{c}{\textbf{CodeLLaMa}} \\ \midrule
                        & Acc. & Pre. & Rec.  & \multicolumn{1}{l|}{F1}   & Acc. & Pre. & Rec. & \multicolumn{1}{l|}{F1}   & Acc      & Pre.    & Rec.    & F1      \\ \midrule
\textbf{Identification} & 0.70 & 0.80 & 0.70  & \multicolumn{1}{l|}{0.67} & 0.56 & 0.57 & 0.56 & \multicolumn{1}{l|}{0.54} & 0.61     & 0.78    & 0.61    & 0.54    \\
CWE-787                 & 0.52 & 0.5  & 0.26  & \multicolumn{1}{l|}{0.34} & 0.23 & 0.5  & 0.11 & \multicolumn{1}{l|}{0.19} & 0.0      & 0.0     & 0.0     & 0.0     \\
CWE-416                 & 1.0  & 1.0  & 1.0   & \multicolumn{1}{l|}{1.0}  & 1.0  & 1.0  & 1.0  & \multicolumn{1}{l|}{1.0}  & 0.0      & 0.0     & 0.0     & 0.0     \\
CWE-20                  & 0.29 & 0.5  & 0.14  & \multicolumn{1}{l|}{0.22} & 0.1  & 0.5  & 0.5  & \multicolumn{1}{l|}{0.09} & 0.94     & 0.5     & 0.47    & 0.48    \\
CWE-125                 & 0.73 & 0.5  & 0.36  & \multicolumn{1}{l|}{0.42} & 0.07 & 0.5  & 0.04 & \multicolumn{1}{l|}{0.07} & 0.0      & 0.0     & 0.0     & 0.0     \\
CWE-476                 & 1.0  & 1.0  & 1.0   & \multicolumn{1}{l|}{1.0}  & 0.58 & 0.5  & 0.29 & \multicolumn{1}{l|}{0.36} & 0.0      & 0.0     & 0.0     & 0.0     \\
CWE-190                 & 0.84 & 0.5  & 0.42  & \multicolumn{1}{l|}{0.45} & 0.5  & 0.5  & 0.25 & \multicolumn{1}{l|}{0.33} & 0.0      & 0.0     & 0.0     & 0.0     \\
CWE-119                 & 0.42 & 0.5  & 0.21  & \multicolumn{1}{l|}{0.29} & 0.36 & 0.5  & 0.18 & \multicolumn{1}{l|}{0.26} & 0.0      & 0.0     & 0.0     & 0.0     \\
CWE-798                 & 0.75 & 0.5  & 0.375 & \multicolumn{1}{l|}{0.42} & 0.4  & 0.5  & 0.2  & \multicolumn{1}{l|}{0.29} & 0.0      & 0.0     & 0.0     & 0.0     \\ \midrule
                        & \multicolumn{4}{c|}{\textbf{Mistral}}           & \multicolumn{4}{c|}{\textbf{LLaMa 3}}          & \multicolumn{4}{c}{\textbf{CodeGen 2}} \\ \midrule
                        & Acc. & Pre. & Rec.  & \multicolumn{1}{l|}{F1}   & Acc. & Pre. & Rec. & \multicolumn{1}{l|}{F1}   & Acc.     & Pre.    & Rec.    & F1      \\ \midrule
\textbf{Identification} & 0.54 & 0.51 & 0.52  & \multicolumn{1}{l|}{0.54} & 0.5  & 0.25 & 0.5  & \multicolumn{1}{l|}{0.33} & 0.54     & 0.60    & 0.51    & 0.52    \\
CWE-787                 & 0.0  & 0.0  & 0.0   & \multicolumn{1}{l|}{0.0}  & 0.84 & 0.5  & 0.47 & \multicolumn{1}{l|}{0.48} & 0.0      & 0.0     & 0.0     & 0.0     \\
CWE-416                 & 1.0  & 1.0  & 1.0   & \multicolumn{1}{l|}{1.0}  & 1.0  & 1.0  & 1.0  & \multicolumn{1}{l|}{1.0}  & 0.0      & 0.0     & 0.0     & 0.0     \\
CWE-20                  & 0.0  & 0.0  & 0.0   & \multicolumn{1}{l|}{0.0}  & 0.94 & 0.5  & 0.47 & \multicolumn{1}{l|}{0.48} & 0.0      & 0.0     & 0.0     & 0.0     \\
CWE-125                 & 0.0  & 0.0  & 0.0   & \multicolumn{1}{l|}{0.0}  & 0.8  & 0.5  & 0.4  & \multicolumn{1}{l|}{0.44} & 0.0      & 0.0     & 0.0     & 0.0     \\
CWE-476                 & 0.0  & 0.0  & 0.0   & \multicolumn{1}{l|}{0.0}  & 0.82 & 0.5  & 0.41 & \multicolumn{1}{l|}{0.45} & 1.0      & 1.0     & 1.0     & 1.0     \\
CWE-190                 & 0.0  & 0.0  & 0.0   & \multicolumn{1}{l|}{0.0}  & 1.0  & 1.0  & 1.0  & \multicolumn{1}{l|}{1.0}  & 0.0      & 0.0     & 0.0     & 0.0     \\
CWE-119                 & 0.41 & 0.32 & 0.57  & \multicolumn{1}{l|}{0.35} & 0.90 & 0.50 & 0.44 & \multicolumn{1}{l|}{0.47} & 0.0      & 0.0     & 0.0     & 0.0     \\
CWE-798                 & 0.0  & 0.0  & 0.0   & \multicolumn{1}{l|}{0.0}  & 0.93 & 0.50 & 0.46 & \multicolumn{1}{l|}{0.47} & 0.0      & 0.0     & 0.0     & 0.0     \\ \bottomrule
\end{tabular}
\label{tab:motivation}
\end{table*}

Compiling an open-source repository is challenging because it requires many software and hardware dependencies to run appropriately and be compiled into binaries listed in the \textit{makefile}. We explored ten popular C repositories from GitHub, mentioned in Table \ref{tab:repositories}. Functions from these repositories were used to generate an adversarial attack on code. After getting the repositories, we extracted the function name from the source code function by parsing the function definition with Tree-Sitter \cite{brunsfeld2018tree} and using an S-expression to extract the function name.

\begin{table}[hb]
\centering
\caption{A brief description of the six different LLMs we tested to analyze their performance on analyzing code vulnerability.}
\begin{tabular}{@{}lllll@{}}
\toprule
\textbf{Model} & \textbf{Src.}     & \textbf{Par.} & \textbf{Modality} & \textbf{Access}  \\ \midrule
GPT-4          & OpenAI              & 1.7T             & Text/Code     & \multirow{2}{*}{API}  \\
Gemini         & Google              & $\sim$           & Text/Code     &                       \\ \midrule
CodeLLaMa      & Meta AI             & 7B               & Code              & \multirow{5}{*}{Open} \\
CodeGen2       & Salesforce & 7B               & Code              &                       \\
LLaMa 3        & Meta AI             & 8B               & Text/Code     &                       \\
Mistral        & Mistral AI          & 7B               & Text/Code     &                       \\ 
\bottomrule
\end{tabular}
\label{tab:test_models}
\end{table}

We randomly selected 200 functions from these repositories and injected vulnerabilities using GPT-4. Each function was appended with instructions on how to inject vulnerability. However, some injected vulnerable code had compilation errors. Therefore, we had to remove some of them, totaling 138 samples. Furthermore, we use the original non-injected function as a non-vulnerable sample in our adversarial dataset, totaling 276 decompiled code samples. Then, we compiled both repositories with the injected vulnerable functions using the GCC compiler on a DGX A100 server with an x86\_64 Linux operating system. Then, we used Ghidra \cite{ghidra} to decompile binaries into decompiled code. 



Although we provided GPT-4 with strict instructions on injecting vulnerabilities without creating potential errors, \texttt{GPT-4} occasionally introduced compiler errors that would prevent the build of the vulnerable repository. Some of these compiler errors included accessing fictitious fields of structures, calling functions that did not exist, and minor syntax errors. Initially, we randomly picked 200 code samples to inject vulnerability. However, out of the 175 samples, 62 samples were not compilable, which we ignored.

Table \ref{tab:cwe_count} shows the total number of decompiled vulnerable functions per CWE category.

\section{Reasoning of Weak Performance in Investigation}

\subsection{Reasoning on Poor Performance on LLMs}
\label{sec:challenges}
Reverse engineers face many challenges when analyzing decompiled code. Understanding the objective and semantics of decompiled code is generally more complex than understanding the semantics of source code. During the decompilation process, the variables or the flow of statements sometimes get changed for optimization. As a result, working with decompiled code for tasks such as vulnerability analysis or decompiled code summarization \cite{al2023extending} is more challenging. Some of the primary reasons for poor performance could be directly related to the removal of comments during decompilation, function name changes to the memory address of the function, complex control flow, and obfuscation of variable names.

\paragraph{\textbf{C1: Comments do not Exist.}} When source code is compiled, the compiler typically ignores comments and no longer exists in the compiled binary. Therefore, comments are irrecoverable in decompiled code. Without comments, comprehending decompiled code is incredibly challenging and time-consuming as it provides limited information about its purpose and intended behavior. Therefore, the reverse engineer has to derive meaning from syntax and structure. 

\paragraph{\textbf{C2: Ambiguous Function Calls.}} When source code is compiled, the compiler may optimize the code by replacing standard function calls, such as \texttt{strcpy}, with a custom function that performs the same task more efficiently. Alternatively, the compiler may optimize the binary by inlining the instructions of common function calls in the callee function, effectively removing the function call altogether. This may prove challenging for reverse engineers attempting to understand code semantics through commonly called functions. This usually happens during decompilation since the dissembler does not know the original function name, and the replaced string \texttt{0x818042DOE} is the function's address in the system memory. 

\begin{table*}[t]
\centering
\caption{Binary Functions across four Computer Architectures and two different Optimization Levels. Our proposed dataset \texttt{DeVulBin} contains 150,872 binary functions, over 40 CWE Classes}
\begin{tabular}{@{}lllllllll@{}}
\toprule
\multicolumn{1}{l|}{\textbf{Architecture}}          & \multicolumn{1}{l|}{\textbf{Optimization}} & \multicolumn{3}{c|}{\textbf{Vulnerable}}                                                                                                                                                                                         & \multicolumn{3}{c|}{\textbf{Non-vulnerable}}                                                                                                                                                                                                                                     & \textbf{Total} \\ \midrule
\multicolumn{1}{l|}{}                               & \multicolumn{1}{l|}{}                      & \multicolumn{1}{c}{\textbf{Vul Func.}} & \multicolumn{1}{c}{\textbf{\begin{tabular}[c]{@{}c@{}}Avg. \\ Token/Line\end{tabular}}} & \multicolumn{1}{c|}{\textbf{\begin{tabular}[c]{@{}c@{}}Avg. Token\\  in Descs.\end{tabular}}} & \multicolumn{1}{c}{\textbf{\begin{tabular}[c]{@{}c@{}}Non-Vul.\\  Func.\end{tabular}}} & \multicolumn{1}{c}{\textbf{\begin{tabular}[c]{@{}c@{}}Avg. \\ Token/Line\end{tabular}}} & \multicolumn{1}{c|}{\textbf{\begin{tabular}[c]{@{}c@{}}Avg. Token\\  in Descs.\end{tabular}}} &                \\ \midrule
\multicolumn{1}{l|}{\textbf{x86}}                   & \multicolumn{1}{l|}{O0}                    & 11924                                  & 73/31                                                                                   & \multicolumn{1}{l|}{167}                                                                      & 23743                                                                                  & 66/29                                                                                   & \multicolumn{1}{l|}{96}                                                                       & 35667          \\
\multicolumn{1}{l|}{}                               & \multicolumn{1}{l|}{O3}                    & 11889                                  & 143/52                                                                                  & \multicolumn{1}{l|}{167}                                                                      & 10345                                                                                  & 151/45                                                                                  & \multicolumn{1}{l|}{95}                                                                       & 22234          \\ \midrule
\multicolumn{1}{l|}{\multirow{2}{*}{\textbf{x64}}}  & \multicolumn{1}{l|}{O0}                    & 5956                                   & 89/35                                                                                   & \multicolumn{1}{l|}{167}                                                                      & 11867                                                                                  & 85/34                                                                                   & \multicolumn{1}{l|}{96}                                                                       & 17823          \\
\multicolumn{1}{l|}{}                               & \multicolumn{1}{l|}{O3}                    & 5962                                   & 157/53                                                                                  & \multicolumn{1}{l|}{167}                                                                      & 7210                                                                                   & 189/54                                                                                  & \multicolumn{1}{l|}{97}                                                                       & 13172          \\ \midrule
\multicolumn{1}{l|}{\multirow{2}{*}{\textbf{ARM}}}  & \multicolumn{1}{l|}{O0}                    & 5962                                   & 83/35                                                                                   & \multicolumn{1}{l|}{167}                                                                      & 11854                                                                                  & 74/32                                                                                   & \multicolumn{1}{l|}{96}                                                                       & 17816          \\
\multicolumn{1}{l|}{}                               & \multicolumn{1}{l|}{O3}                    & 5960                                   & 123/48                                                                                  & \multicolumn{1}{l|}{167}                                                                      & 7210                                                                                   & 163/52                                                                                  & \multicolumn{1}{l|}{97}                                                                       & 13170          \\ \midrule
\multicolumn{1}{l|}{\multirow{2}{*}{\textbf{MIPS}}} & \multicolumn{1}{l|}{O0}                    & 5960                                   & 81/32                                                                                   & \multicolumn{1}{l|}{167}                                                                      & 11866                                                                                  & 73/30                                                                                   & \multicolumn{1}{l|}{96}                                                                       & 17826          \\
\multicolumn{1}{l|}{}                               & \multicolumn{1}{l|}{O3}                    & 5959                                   & 132/45                                                                                  & \multicolumn{1}{l|}{167}                                                                      & 7205                                                                                   & 204/57                                                                                  & \multicolumn{1}{l|}{97}                                                                       & 13164          \\ \midrule
\multicolumn{9}{r}{\textbf{40 Different Categories of CWEs Totaling of 150872}}                                                                                                                                                                                                                                                                                                                                                                                                                                                                                                                                                         \\ \bottomrule
\end{tabular}
\label{tab:metadata}
\end{table*}

\paragraph{\textbf{C3: Complex Control Flow.}} The control flow of the source code may be optimized and modified in the compiled binary. As a result, the decompiled code may have a more complex control flow of statements, which is difficult to understand from a reverse engineer's perspective. A common example is loop unrolling \cite{sarkar2000optimized, huang1999generalized}, where a loop is unraveled as a sequence of instructions instead of jumping back to the beginning of the loop until a condition is met. However, these optimizations can sometimes be unusual and confusing when comprehending the flow of decompiled code. 


\section{DeBinVul Dataset Preparation}

In this section, we highlight the technical details of how we extract each dataset component in detail.

\textbf{Function Extraction.}
To extract all function definitions from the file, we used the S-expression of Tree-Sitter. In Tree-sitter, an S-expression (symbolic expression) is a way to represent the Abstract Syntax Tree (AST) of source code in a nested, parenthetical format. Each node in the tree is represented as a list, starting with the node type followed by its children, which can be terminal tokens or further nested lists. This format provides a clear and concise textual representation of the code's syntactic structure. To extract the functions, we used the following S-expression, $(function\_definition) @func-def$.

After extracting the functions using S-expression, our next task is to separate vulnerable from non-vulnerable functions. We found a specific pattern that makes this task more straightforward for us. We observe that all function definitions that are extracted from the file either contain "good" (a benign entry) or "bad" (a vulnerable entry) in the function's name. For each of the extracted function definitions, we used another S-expression to extract function names from each function definition: $(function\_declarator (identifier) @func\_name)$. Complete definitions of these S-expressions are available in the repository we provided earlier.

After extracting the functions and function names, our next task is to classify the functions. This part is relatively straightforward. If the function name contains the substring "good," we consider it a benign or non-vulnerable function. However, if the function contains the sub-string "bad," we consider the function vulnerable. If the function is non-vulnerable, the function name has the format that appears as \textit{CWEXXX\_rest\_of\_function\_name}. Therefore, we take the first part of the function name (\textit{CWEXXX}) to capture the CWE number of the vulnerable code. Table \ref{tab:cwe_count} shows the total number of decompiled functions we generated for each CWE.

\textbf{Compilation and Optimization.}
Our compilation process strategically employs the $-O0$ and $-O3$ optimization levels to assess the impact of compiler optimizations on the security and functionality of executables. By selecting these two extreme levels, we can thoroughly evaluate how optimizations influence executable behavior in ways that intermediate levels, $-O1$ and $-O2$, may not fully reveal. The $-O0$ level, which applies no optimization, ensures that the compiled code remains a straightforward representation of the source code. This direct correspondence is critical for accurately tracing vulnerabilities back to their source, providing a clear baseline for understanding the application's intended behavior.

In contrast, the $-O3$ level introduces aggressive optimizations such as function inlining, loop unrolling, and advanced vectorization. These can enhance performance and efficiency and potentially introduce or expose vulnerabilities, such as buffer overflows, specifically related to these optimization techniques. Moreover, $-O3$ mimics the high-performance conditions often found in production environments, making it invaluable for simulating real-world application scenarios. This dual approach, employing both $-O0$ and $-O3$, allows us to capture a comprehensive range of effects— from no optimization to maximum optimization—thereby providing a broad spectrum analysis of how different optimization levels can affect an executable’s performance, size, and, crucially, its security properties. This method ensures we identify any vulnerabilities introduced or masked by compiler optimizations, offering a robust evaluation that intermediate optimization levels might overlook.

We utilized the following compiler commands: $gcc -O0 (x86)$, $gcc -O3 (x86)$, $clang -O0 (x86)$, $clang -O3 (x86)$, $aarch64-linux-gnu-gcc -O0$ (ARM), and $aarch64-linux-gnu-gcc -O3$ (ARM). The $-D INCLUDEMAIN$ option was included to define the main function necessary for compiling the CWE code. We compiled the source code twice for each compiler command using the $-D OMITGOOD$ and $-D OMITBAD$ options to generate the vulnerable and benign executables respectively. This systematic approach ensured that we could thoroughly examine the impact of different compilers, optimization levels, and code variants on the security properties of the executables.

Our study assesses the impact of compiler optimizations on security and functionality by using the $-O0$ and $-O3$ optimization levels. The $-O0$ level, with no optimizations, provides a direct correspondence to the source code, which is essential for tracing vulnerabilities accurately. Conversely, the $-O3$ level applies aggressive optimizations that can enhance performance but also introduce or expose vulnerabilities, simulating high-performance production environments. This dual approach allows us to capture a wide range of effects, providing a thorough analysis of how different optimization levels influence executable behavior. By comparing the extremes of no optimization and maximum optimization, we ensure a robust evaluation that intermediate levels might miss.

\begin{table}[ht]
\centering
\caption{CWE Count Per Class of Decompiled Binary Code}
\begin{tabular}{@{}llll@{}}
\toprule
\textbf{CWE Number} & \textbf{Count} & \textbf{CWE Number} & \textbf{Count} \\ \midrule
CWE-127             & 3344           & CWE-666             & 799            \\
CWE-590             & 3330           & CWE-510             & 699            \\
CWE-124             & 3340           & CWE-426             & 660            \\
CWE-121             & 3340           & CWE-467             & 479            \\
CWE-401             & 3339           & CWE-415             & 370            \\
CWE-122             & 3339           & CWE-506             & 340            \\
CWE-761             & 3339           & CWE-475             & 320            \\
CWE-690             & 3339           & CWE-319             & 319            \\
CWE-126             & 3338           & CWE-464             & 300            \\
CWE-427             & 3338           & CWE-459             & 260            \\
CWE78               & 3338           & CWE-773             & 230            \\
CWE-789             & 3338           & CWE-476             & 180            \\
CWE-606             & 3338           & CWE-605             & 180            \\
CWE-680             & 2797           & CWE-469             & 160            \\
CWE-665             & 1696           & CWE-675             & 140            \\
CWE-758             & 1677           & CWE-404             & 100            \\
CWE-123             & 1349           & CWE-775             & 70             \\
CWE-617             & 1087           & CWE-681             & 70             \\
CWE-457             & 980            & CWE-688             & 40             \\
CWE-416             & 830            & CWE-685             & 30             \\ \bottomrule
\end{tabular}
\label{tab:cwe_count}
\end{table}

\textbf{Decompilation}
During decompilation, we use the extra flag $-s$. The $-s$ flag in GCC instructs the compiler to strip symbol information from the resulting executable, including function names. This significantly reduces the executable's size but also hinders debugging and reverse engineering efforts. Stripped binaries can be more challenging to analyze and understand, potentially making it more difficult for attackers to exploit vulnerabilities. However, it's essential to note that while -s removes human-readable function names, it doesn't obscure the underlying code logic or prevent advanced reverse engineering techniques.

\textbf{Instruction Generation}
We created 80 instructions in total, and 20 instruction
We created 20 instructions for each task, totaling 80 instructions for each task. We used four specially curated prompts using GPT-4 to automatically generate the 80 instructions for all of the four tasks in decompiled binary analysis.

\begin{table*}[ht]
\centering
\caption{The instructions we used to investigate LLMs to determine their performance on decompiled code vulnerability identification and classification tasks.}
\begin{tabular}{l|l|l}
\hline
\textbf{GPT-4 Instruction}                                                                                                                                                                                                                                                                                                                                                                                                                                                                                                                                                                       & \textbf{Gemini Instruction}                                                                                                                                                                                                                                                                                                                                                                                                                                                                                                                                                                                                                                      & \textbf{Other LLM Instruction}                                                                                                                                                                                                                                                                                                                                                                                                                                                                                                                                                                                                                                                                               \\ \hline
\begin{tabular}[c]{@{}l@{}}You are an expert at analyzing software \\ security vulnerabilities in static source\\ code and decompiled binaries. \\ Task Description: \\ Task 1: Your task is to detect if a\\ vulnerability exists in the following \\ code. Answer with\\ "Yes/No" only if a vulnerability \\ exists. \\ Task 2: If your answer is "Yes" to \\ detecting  vulnerability, print only\\  the CWE number \\ of the vulnerability. Do not produce \\ any other output more than you are \\ asked. Do not use any context from \\ any previous input or output.\\ Code:\end{tabular} & \begin{tabular}[c]{@{}l@{}}You are an expert at analyzing software \\ security vulnerabilities in static source \\ code and decompiled binaries. \\ Task Description: Task 1: Your task is to\\ detect if a vulnerability exists in the \\ following code. Answer with "Yes/No"\\ only if vulnerability exists; Task 2: If \\ your answer is "Yes" to detecting \\ vulnerability, print only the CWE \\ number of the vulnerability. Do not \\ produce any other output more than \\ you are asked. Do not use any context \\ from any previous input or output. \\ Answer only in the following format, no\\ extra link: "Yes/No CWE-XXX" \\ Code:\end{tabular} & \begin{tabular}[c]{@{}l@{}}You are an expert at analyzing software\\ security vulnerabilities in static source \\ code and decompiled binaries. \\ Task Description: \\ Task 1: Your task is to detect if a \\ vulnerability exists in the following \\ code. Answer with "Yes/No" only \\ if a vulnerability exists. Task 2: \\ If your answer is "Yes" to detecting \\ vulnerability, print only the CWE\\ number of the vulnerability. Do \\ not produce any other output more\\ than you are asked. Do not use any \\ context from any previous input or \\ output. Answer only in the following\\ format; do not describe the CWE \\ number and no \\ extra link: "Yes/No CWE-XXX"\\ Code:\end{tabular} \\ \hline
\end{tabular}
\label{tab:insts}
\end{table*}

\textbf{Post-processing.}
Following the extraction, compilation, and decompilation of functions using Ghidra, several post-processing steps were undertaken to ensure the dataset's quality and suitability for vulnerability analysis. First, instances of gibberish code that would not compile were identified and removed to prevent skewed results. Additionally, empty methods containing only method names without executable code were eliminated, as they provided no value to the analysis. We also encountered numerous codes that were either identical or too similar, differing only in variable names (e.g., `srini\_string` versus `string\_srini`); these redundancies were systematically removed to maintain dataset diversity and prevent bias. Furthermore, many CWE categories lacked viable code examples due to insufficient training data in the original CWE dataset. To address this, we employed synthetic code generation and semi-supervised learning techniques to augment the dataset, thereby increasing the representation of underrepresented CWEs. These post-processing steps were crucial in refining the dataset, ensuring it was robust, diverse, and ready for accurate vulnerability analysis. Table \ref{tab:metadata} briefly overviews our proposed dataset.

\begin{table*}[]
\centering
\caption{RQ1: Comparison of Classification on 20 CWEs against various LLMs}
\begin{tabular}{@{}llllllllllll@{}}
\toprule
                                                              &                 & \multicolumn{10}{c}{\textbf{CWE}}                                                                                                                                                                                                                                                                                                                           \\ \midrule
\textbf{Model}                                                & \textbf{Metric} & \textbf{401}                     & \textbf{690}                     & \textbf{124}                     & \textbf{761}                     & \textbf{590}                     & \textbf{121}                     & \textbf{789}                     & \textbf{122}                     & \textbf{606}                     & \textbf{127}                     \\ \midrule
\multirow{3}{*}{CodeLLaMa} & Pre.            & 0.83                             & 0.9                              & 0.94                             & 0.97                             & 0.93                             & 1                                & 0.68                             & 0.59                             & 0.71                             & 0.93                             \\
                                                              & Rec             & 0.51                             & 0.79                             & 1                                & 0.9                              & 0.9                              & 1                                & 0.48                             & 0.96                             & 0.52                             & 0.61                             \\
                                                              & F1              & 0.63                             & 0.84                             & 0.97                             & 0.93                             & 0.91                             & 1                                & 0.57                             & 0.73                             & 0.6                              & 0.74                             \\ \midrule
\multirow{3}{*}{CodeGen2}                                     & Pre.            & 0.98                             & 0.98                             & 0.81                             & 1                                & 0.86                             & 0.86                             & 0.92                             & 1                                & 0.79                             & 0.59                             \\
                                                              & Rec             & 1                                & 1                                & 0.62                             & 1                                & 0.97                             & 1                                & 1                                & 0.91                             & 1                                & 0.45                             \\
                                                              & F1              & 0.99                             & 0.99                             & 0.7                              & 1                                & 0.91                             & 0.93                             & 0.96                             & 0.95                             & 0.88                             & 0.51                             \\ \midrule
\multirow{3}{*}{Mistral}                                      & Pre.            & 0.84                             & 0.89                             & 0.94                             & 0.97                             & 1                                & 0.91                             & 0.75                             & 0.74                             & 0.73                             & 1                                \\
                                                              & Rec             & 0.57                             & 0.97                             & 1                                & 1                                & 0.83                             & 1                                & 0.44                             & 0.96                             & 0.48                             & 0.39                             \\
                                                              & F1              & 0.68                             & 0.93                             & 0.97                             & 0.98                             & 0.91                             & 0.95                             & 0.56                             & 0.84                             & 0.58                             & 0.56                             \\ \midrule
\multirow{3}{*}{LLaMa 3}                                      & Pre.            & 0.9                              & 0.94                             & 0.97                             & 0.89                             & 1                                & 1                                & 0.76                             & 0.69                             & 0.83                             & 0.93                             \\
                                                              & Rec             & 0.76                             & 1                                & 1                                & 1                                & 0.9                              & 1                                & 0.81                             & 1                                & 0.83                             & 0.61                             \\
                                                              & F1              & 0.82                             & 0.97                             & 0.99                             & 0.94                             & 0.95                             & 1                                & 0.79                             & 0.82                             & 0.83                             & 0.74                             \\ \midrule
\multirow{3}{*}{StarCoder}                                    & Pre.            & 0.78                             & 0.69                             & 0.87                             & 0.94                             & 1                                & 0.97                             & 0.79                             & 0.93                             & 0.77                             & 0.92                             \\
                                                              & Rec             & 0.57                             & 0.85                             & 1                                & 1                                & 0.9                              & 1                                & 0.41                             & 1                                & 0.74                             & 0.52                             \\
                                                              & F1              & 0.66                             & 0.76                             & 0.93                             & 0.97                             & 0.95                             & 0.98                             & 0.54                             & 0.96                             & 0.76                             & 0.67                             \\ \midrule
                                                              &                 & \multicolumn{1}{r}{\textbf{122}} & \multicolumn{1}{r}{\textbf{121}} & \multicolumn{1}{r}{\textbf{789}} & \multicolumn{1}{r}{\textbf{680}} & \multicolumn{1}{r}{\textbf{758}} & \multicolumn{1}{r}{\textbf{416}} & \multicolumn{1}{r}{\textbf{665}} & \multicolumn{1}{r}{\textbf{457}} & \multicolumn{1}{r}{\textbf{123}} & \multicolumn{1}{r}{\textbf{617}} \\ \midrule
\multirow{3}{*}{CodeLLaMa} & Pre.            & 0.9                              & 0.51                             & 1                                & 0.78                             & 0.94                             & 1                                & 0.62                             & 0.73                             & \textbf{1}                       & \textbf{0.75}                    \\
                                                              & Rec             & 0.86                             & 1                                & 0.81                             & 0.9                              & 0.89                             & 0.87                             & 0.71                             & 0.92                             & 0.73                             & 0.9                              \\
                                                              & F1              & 0.88                             & 0.68                             & 0.89                             & 0.84                             & 0.91                             & 0.93                             & 0.67                             & 0.81                             & 0.84                             & 0.82                             \\ \midrule
\multirow{3}{*}{CodeGen2}                                     & Pre.            & 0.95                             & 0.95                             & 0.94                             & 1                                & 0.92                             & 0.76                             & 0.7                              & 0.8                              & 1                                & 0.57                             \\
                                                              & Rec             & 0.86                             & 0.9                              & 0.83                             & 0.88                             & 0.71                             & 1                                & 0.93                             & 0.92                             & 0.83                             & 0.44                             \\
                                                              & F1              & 0.9                              & 0.92                             & 0.88                             & 0.94                             & 0.8                              & 0.86                             & 0.8                              & 0.86                             & 0.91                             & 0.5                              \\ \midrule
\multirow{3}{*}{Mistral}                                      & Pre.            & 0.83                             & 0.49                             & 1                                & 0.83                             & 0.94                             & 0.82                             & 0.68                             & 0.79                             & 1                                & 0.89                             \\
                                                              & Rec             & 0.91                             & 1                                & 1                                & 0.95                             & 0.94                             & 0.93                             & 0.93                             & 0.92                             & 1                                & 0.8                              \\
                                                              & F1              & 0.87                             & 0.66                             & 1                                & 0.88                             & 0.94                             & 0.88                             & 0.79                             & 0.85                             & 1                                & 0.84                             \\ \midrule
\multirow{3}{*}{LLaMa 3}                                      & Pre.            & 0.91                             & 0.84                             & 0.95                             & 0.95                             & 0.95                             & 1                                & 0.87                             & 0.92                             & 1                                & 0.82                             \\
                                                              & Rec             & 0.91                             & 1                                & 1                                & 0.9                              & 1                                & 0.93                             & 0.93                             & 0.92                             & 1                                & 0.9                              \\
                                                              & F1              & 0.91                             & 0.91                             & 0.98                             & 0.92                             & 0.97                             & 0.97                             & 0.9                              & 0.92                             & 1                                & 0.86                             \\ \midrule
\multirow{3}{*}{StarCoder}                                    & Pre.            & 0.87                             & 0.5                              & 0.78                             & 0.95                             & 0.93                             & 1                                & 0.86                             & 0.73                             & 1                                & 1                                \\
                                                              & Rec             & 0.91                             & 1                                & 1                                & 1                                & 0.78                             & 0.93                             & 0.86                             & 0.92                             & 0.82                             & 0.8                              \\
                                                              & F1              & 0.89                             & 0.67                             & 0.88                             & 0.98                             & 0.85                             & 0.97                             & 0.86                             & 0.81                             & 0.9                              & 0.89                             \\ \bottomrule
\end{tabular}
\label{tab:base_clsssification}
\end{table*}

\begin{table*}[t]
\centering
\caption{Prompts used as an input for GPT-4 to generate 20 Instructions for each task, namely, i) Identification, ii) Classification, iii) Function Name Prediction and iv) Description}
\begin{tabular}{ll}
\hline
\textbf{Task}                                                           & \textbf{Prompt}                                                                                                                                                                                                                                                                                                                                                                                                                                                                                                                                                                                                                                                                                                                                                                                                                                                                                                                                                                                                                                                                                                                                                                                                                                                                                                                         \\ \hline
\textbf{Ident.}                                                         & \begin{tabular}[c]{@{}l@{}}Consider yourself as a Code Security Reverse engineer Specialist. \\ Task: You are to generate a prompt to ask a large language model to Identify Vulnerabilities in decompiled\\ code. Make sure that the prompt ensures the answer is in "Yes/No" format and that no extra content is \\ generated by the LLM when using the prompt. \\ Now, create 20 different variants of these prompts. \\ Here are two sample examples: \\ 1. "You are a Reverse Engineer of Code. Detect the presence of events of vulnerabilities that exist in the\\ given code. \\ If a vulnerability exists, answer "YES", otherwise, answer "NO". Do not produce any extra outputs. \\ 2. "As an event Vulnerability expert of decompiled code, analyze the code to find whether any \\ vulnerability exists in the following code. \\ If you think the code has vulnerability, produce a single word "YES", and if vulnerability does not exist, \\ reply "NO" only. Do not output any extra words or URLs.\end{tabular}                                                                                                                                                                                                                                                                                              \\ \hline
\textbf{Class.}                                                         & \begin{tabular}[c]{@{}l@{}}Consider yourself as a Code Security Reverse engineer Specialist. \\ Task: You are to generate a prompt to ask a large language model to Classify Vulnerabilities in decompiled \\ code. The classification is based on the CWE numbers. Make sure that the prompt ensures the answer is in\\ "CWE-XXX" \\ format and that no extra content is generated by the LLM when using the prompt. Here "XXX" refers to \\ the vulnerability number. \\ Now, create 20 different variants of these prompts. \\ Here are two sample examples: \\ 1. "You are a Reverse Software Engineer of Decompiled Code. Classify the presence of vulnerabilities\\ that exist \\ in the given code. \\ Generate the vulnerability classification in the following format only "CWE-XXX". Do not produce \\ any extra outputs. \\ 2. "As an expert classifying Vulnerability in decompiled code, analyze the code to find the \\ vulnerability category existing in the following code. Print only the CWE number of the vulnerability \\ in this format "CWE-XXX". \\ Do not output any extra words or URLs.\end{tabular}                                                                                                                                                                                               \\ \hline
\begin{tabular}[c]{@{}l@{}}\textbf{Func.} \\ \textbf{Name} \\ \textbf{Pred.}\end{tabular} & \begin{tabular}[c]{@{}l@{}}Consider yourself as a Code Security Reverse engineer Specialist who can predict function names on \\ decompiled binary code. \\ Task: You are to generate a prompt to ask a large language model to predict function name in decompiled code.\\ The function name can only have the characters supported in C/C++ programming language, and the outcome\\ would only consist of a single word, and no extra content is generated by the LLM when using the prompt. \\ Now, create 20 different variants of these prompts. \\ Here are two sample examples: \\ 1. "You are a Reverse Software Engineer of Decopiled Code. Predict the name of the decompiled code. \\ Generate the function name only in a single word. You can use camelCasing or Snake\_Casing. \\ Do not produce any extra outputs. \\ 2. "As an expert classifying Vulnerability in decompiled code, analyze the code to determine\\ function name in the following code. \\ Print only the Function name in a single word. You are allowed to use snake\_casing or camelCasing \\ to generate the function names. Do not output any extra words or URLs.\end{tabular}                                                                                                                                                          \\ \hline
\textbf{Desc.}                                                          & \begin{tabular}[c]{@{}l@{}}Consider yourself as a Code Security Reverse engineer Specialist who can generate the description of a \\ decompiled code. \\ Task: You are to generate a prompt to ask a large language model to Describe the objective\\ and/or Vulnerabilities in decompiled code. The description should explain the flow of the code but not \\ specify any variable or function names. The descriptions should be generic enough to be used in a\\ decompiled code as well. \\ Now, create 20 different variants of these prompts. \\ Here are two sample examples: \\ 1. "You are a Reverse Software Engineer of Decopiled Code. Describe the objective and the security \\ issues that exist in the given code. Make sure you generate a generic description of the vulnerability \\ without specifying function or variable names. Please ensure that the generated description \\ can be used in decompiled code. \\ 2. "As an expert explaining the objectives and vulnerability in decompiled code, analyze the code \\ to explain the objective and vulnerability. Ensure the generalizability of the description by not \\ mentioning the function and the variable names as the description will be used in a decompiled \\ code where the variable and function names are obfuscated."\end{tabular} \\ \hline
\end{tabular}
\label{tab:prompt}
\end{table*}

\begin{table*}[t]
\centering
\caption{Average F1 Score Comparison on Base vs. Trained LLMs for Classification Task.}
\begin{tabular}{@{}lllllllllll@{}}
\toprule
                 & \multicolumn{2}{l}{\textbf{CodeLLaMa}} & \multicolumn{2}{l}{\textbf{CodeGen2}} & \multicolumn{2}{l}{\textbf{Mistral}} & \multicolumn{2}{l}{\textbf{LLaMa 3}} & \multicolumn{2}{l}{\textbf{StarCoder}} \\ \midrule
                 & \textbf{Base}    & \textbf{Trained}    & \textbf{Base}    & \textbf{Trained}   & \textbf{Base}   & \textbf{Trained}   & \textbf{Base}   & \textbf{Trained}   & \textbf{Base}    & \textbf{Trained}    \\
\textbf{CWE-124} & 0                & 0.63                & 0                & 0.99               & 0               & 0.68               & 0               & 0.82               & 0                & 0.66                \\
\textbf{CWE-427} & 0                & 0.84                & 0                & 0.99               & 0               & 0.93               & 0               & 0.97               & 0                & 0.76                \\
\textbf{CWE-401} & 0.11             & 0.97                & 0                & 0.7                & 0               & 0.97               & 0               & 0.99               & 0.11             & 0.93                \\
\textbf{CWE-761} & 0.12             & 0.93                & 0                & 1                  & 0               & 0.98               & 0               & 0.94               & 0                & 0.97                \\
\textbf{CWE-590} & 0                & 0.91                & 0                & 0.91               & 0               & 0.91               & 0               & 0.95               & 0                & 0.95                \\
\textbf{CWE-690} & 0.13             & 1                   & 0                & 0.93               & 0               & 0.95               & 0               & 1                  & 0.13             & 0.98                \\
\textbf{CWE-127} & 0                & 0.57                & 0                & 0.96               & 0               & 0.56               & 0               & 0.79               & 0                & 0.54                \\
\textbf{CWE-606} & 0                & 0.73                & 0                & 0.95               & 0               & 0.84               & 0               & 0.82               & 0                & 0.96                \\
\textbf{CWE-126} & 0                & 0.6                 & 0                & 0.88               & 0               & 0.58               & 0               & 0.83               & 0                & 0.76                \\
\textbf{CWE-78}  & 0.16             & 0.74                & 0                & 0.51               & 0               & 0.56               & 0               & 0.74               & 0.16             & 0.67                \\
\textbf{CWE-122} & 0.09             & 0.88                & 0                & 0.9                & 0               & 0.87               & 0               & 0.91               & 0.08             & 0.89                \\
\textbf{CWE-121} & 0.07             & 0.68                & 0                & 0.92               & 0               & 0.66               & 0               & 0.91               & 0.08             & 0.67                \\
\textbf{CWE-789} & 0.09             & 0.89                & 0                & 0.88               & 0               & 1                  & 0               & 0.98               & 0                & 0.88                \\
\textbf{CWE-680} & 0                & 0.84                & 0                & 0.94               & 0               & 0.88               & 0               & 0.92               & 0.1              & 0.98                \\
\textbf{CWE-758} & 0                & 0.91                & 0                & 0.8                & 0               & 0.94               & 0               & 0.97               & 0                & 0.85                \\
\textbf{CWE-416} & 0                & 0.93                & 0                & 0.86               & 0               & 0.88               & 0               & 0.97               & 0                & 0.97                \\
\textbf{CWE-665} & 0                & 0.67                & 0                & 0.8                & 0               & 0.79               & 0               & 0.9                & 0                & 0.86                \\
\textbf{CWE-457} & 0                & 0.81                & 0                & 0.86               & 0               & 0.85               & 0               & 0.92               & 0                & 0.81                \\
\textbf{CWE-123} & 0                & 0.84                & 0                & 0.91               & 0               & 1                  & 0               & 1                  & 0                & 0.9                 \\
\textbf{CWE-617} & 0                & 0.82                & 0                & 0.5                & 0               & 0.84               & 0               & 0.86               & 0                & 0.89                \\
\textbf{Average} & 0.0385           & 0.8095              & 0                & 0.8595             & 0               & 0.8335             & 0               & 0.9095             & 0.033            & 0.844               \\ \bottomrule
\end{tabular}
\label{tab:classification_details}
\end{table*}

\section{Evaluation Metrics}
\paragraph{\textbf{BLEU Score.}}The BLEU \cite{papineni2002bleu} score is a syntax-based way of evaluating machine-generated text between 0 and 1. It is a reference-based metric and may not capture all aspects of translation quality, such as fluency or semantic accuracy.

\paragraph{\textbf{Rouge-L.}}Similar to BLEU, Rouge-L \cite{lin2004rouge} score is also a number between 0 and 1 to measure the syntax-based similarity of two generated texts. It generates a score by quantifying precision and recall by examining the longest common subsequence (LCS) between the generated and reference codes.

\paragraph{\textbf{BERTScore.}} Furthermore, we use BERTScore \cite{zhang2019bertscore} for semantic comparison using cosine similarity score to identify how the generated token matches the ground truth tokens. BERTScore generates an embedding vector for each generated token, performs a cosine similarity with all the ground truth tokens, and averages it to generate the final result. It is a token-based similarity representation vector that represents tokens that permit generating a soft similarity measure instead of exact matching since secure code can be generated in various ways.

\paragraph{\textbf{Cosine Similarity.}} BLEU, Rouge-L, BERTScore compares n-gram-based exact token matching or embedding-based token matching. However, the exact meaning can be kept intact for natural language while the description is written differently. Thus, we aim to measure the semantic similarity between the embedding of the LLM generated text and the ground truth. Formally given the set of sequences of tokens generated by LLM description $D = \{w_1, w_2, ..., w_n\}$ and the ground truth reference description $\hat{D} = \{\hat{w_1}, \hat{w_2}, ..., \hat{w_n}\}$. Then we used the sentence encoder $E$  to generate the embedding $E(D) = \{e_{w_1}, e_{w_2}, ..., e_{w_n}\}$ and  $E(D) = \{e_{\hat{w_1}}, e_{\hat{w_2}}, ..., e_{\hat{w_n}}\}$. Thus the entire semantics of $D$ and $\hat{D}$ are represented by:
\begin{equation}
    e_D = \frac{1}{|D|} \sum_{i=1}^{m} e_{w_i}, e_{\hat{D}} = \frac{1}{|\hat{D}|} \sum_{j=1}^{n} e_{w_j}
\end{equation}

Therefore, we calculate the similarity score as, 

\begin{equation}
    sim(D, \hat{D)} = \frac{e_D \cdot e_{\hat{D}}^T}{||e_D|| \cdot ||e_{\hat{D}}||}
\end{equation}

\section{Further Discussion on RQ1.} 
For the classification task, from Table \ref{tab:base_clsssification}, we can see that CodeGen2 and Starcoder show the highest performance across most CWEs, with an F1 score over 90\% for CWE-590, CWE-121, and CWe-122. Furthermore, Mistral also shows similar performance across all other CWEs. However, CodeLLaMa shows relatively poorer performance. However, we also notice that some CWEs like CWE-121, CodeLLaMa, and LLaMa 3 show a 100\% accurate performance. Moreover, Mistral \cite{jiang2023mistral} shows a 100\% accuracy for CWE-789 and CWE-123, which further elucidates that some models have a slight bias in classifying different CWEs.

We use the same models trained for decompiled code function name prediction and function description generation. From Table \ref{tab:base_name_desc}, we see that LLaMa 3 has the highest Function Name prediction performance across all other models with a BERTScore F1 of 0.97 and a Cosine Similarity score of 0.83. However, for the description generation task, we can see that CodeLLaMa, Mistral, and LLaMa 3 perform superior compared to all other models,  where Mistral and LLaMa show an improvement of 7\% over StarCoder and merely 1\% improvement on CodeLLaMa and CodeGen2.

    \label{9_appendix}

\end{document}